%% file: main.tex
\pdfoutput=1
\documentclass[sigconf,screen]{acmart} 

\copyrightyear{2026}
\acmYear{2026}
\setcopyright{cc}
\setcctype{by}
\acmConference[ICSE '26]{2026 IEEE/ACM 48th International Conference on Software Engineering}{April 12--18, 2026}{Rio de Janeiro, Brazil}
\acmBooktitle{2026 IEEE/ACM 48th International Conference on Software Engineering (ICSE '26), April 12--18, 2026, Rio de Janeiro, Brazil}
\acmPrice{}
\acmDOI{10.1145/3744916.3787800}
\acmISBN{979-8-4007-2025-3/2026/04}

\input{macros}

\usepackage{algorithm}
\begin{document}

\title[Testing Deep Learning Libraries via Neurosymbolic Constraint Learning]{Testing Deep Learning Libraries via \\Neurosymbolic Constraint Learning}

\author{M M Abid Naziri}\authornote{Equal Contribution.}
\affiliation{
  \institution{North Carolina State University}
  \city{Raleigh}
  \state{North Carolina}
  \country{USA}
}
\email{mnaziri@ncsu.edu}

\author{Shinhae Kim}\authornotemark[1]
\affiliation{
  \institution{Cornell University}
  \city{Ithaca}
  \state{New York}
  \country{USA}
}
\email{sk3364@cornell.edu}

\author{Feiran (Alex) Qin}
\affiliation{
  \institution{North Carolina State University}
  \city{Raleigh}
  \state{North Carolina}
  \country{USA}
}
\email{fqin2@ncsu.edu}

\author{Marcelo d'Amorim}
\affiliation{
  \institution{North Carolina State University}
  \city{Raleigh}
  \state{North Carolina}
  \country{USA}
}
\email{mdamori@ncsu.edu}

\author{Saikat Dutta}
\affiliation{
  \institution{Cornell University}
  \city{Ithaca}
  \state{New York}
  \country{USA}
}
\email{saikatd@cornell.edu}

\begin{abstract}
Deep Learning (DL) libraries (\eg{}, \torch) are popular in the
development of AI applications. These libraries are complex and
contain bugs. Researchers have proposed various bug-finding techniques
for such libraries. Yet, there is much room for improvement.
A key challenge in testing DL libraries is the lack of API
specifications. Prior testing approaches often inaccurately model the
input specifications of DL APIs, resulting in missed valid inputs
that could reveal bugs or false alarms due to invalid inputs.




To address this challenge, we develop \tname -- the \emph{first} neurosymbolic
technique to test \dl\ library APIs using dynamically learned input \invariants.
\tname{} leverages the key idea that formal API constraints can be
\emph{learned} from a small number of \NA{automatically generated} seed inputs,
and that the learned constraints can be solved using SMT solvers to generate
valid and diverse test inputs \NA{to test} the API. 

We develop a novel grammar that represents first-order logic formulae over API
parameters and expresses tensor-related properties (e.g., shape, tensor data
types, etc.) as well as relational properties between parameters. We use the
grammar to guide a Large Language Model (LLM) to enumerate syntactically correct
candidate rules, which we then validate using the seed inputs. \NA{Further, we
develop a custom refinement strategy to prune the set of learned rules to
eliminate spurious or redundant rules.}
We use the learned constraints to systematically generate valid and diverse inputs for the
API \NA{by integrating SMT-based solving with randomized sampling.}

We evaluate \tname for testing PyTorch and TensorFlow.
Our results show that \tname's constraints have a recall 
of 94.0\% and a precision of 94.0\% on average.
In terms of coverage, \tname covers \moreCovThanTitanfuzz{},
\moreCovThanACETest{}, and \moreCovThanPathfinder{} more branches than
\titanfuzz, \acetest and \pathfinder, respectively. Using \tname, we also detect
\totalBugs\ new bugs in PyTorch and TensorFlow, \totalBugsConfirmed\ of which
are confirmed.

\end{abstract}  
  
\maketitle

\input{intro}

\input{example}
\input{technique}
\input{exp-setup}
\input{evaluation}
\input{discussion}
\input{validity} 
\input{related}
\input{conclusion}
\begin{acks}
\label{sec:acks}
This work is partially supported by the United States National Science
Foundation (NSF) under Grant Nos. CCF-2349961 and CCF-2319472. We thank Google
for the Google Cloud Platform credits and Meta for the Meta LLM Evaluation
Research Grant. We also thank the anonymous reviewers for their valuable
feedback.
\end{acks}

\bibliographystyle{ACM-Reference-Format}
\bibliography{main}


\end{document}

%% file: macros.tex
\newcommand{\added}[1]{#1}
\newcommand{\deleted}[1]{} 
\newcommand{\removeRow}[1]{}

\setcitestyle{numbers,sort&compress}

\usepackage[table]{xcolor}
\usepackage{colortbl}
\usepackage{mathtools}
\usepackage{multirow}
\usepackage{enumitem}
\usepackage{color}
\usepackage{listings}

\definecolor{pblue}{rgb}{0.13,0.13,1}
\definecolor{pgreen}{rgb}{0,0.5,0}
\definecolor{pred}{rgb}{0.9,0,0}
\definecolor{pgrey}{rgb}{0.46,0.45,0.48}

\lstdefinestyle{custompython}{
    language=Python,
    basicstyle=\ttfamily\footnotesize, 
    commentstyle=\color{pgreen},
    keywordstyle=\color{pblue},
    stringstyle=\color{pred},
    showstringspaces=false,
    breaklines=true,
    numbers=left,
    numberstyle=\tiny\color{pgrey},
    stepnumber=1,
    numbersep=6pt,
    tabsize=2,
    frame=single,              
    rulecolor=\color{black},
    captionpos=b,              
    xleftmargin=2.5em,
    linewidth=0.99\columnwidth 
}

\usepackage{breakurl}
\usepackage{xspace}
\usepackage[round-mode=places,round-precision=1,group-separator={,},group-minimum-digits={3},output-decimal-marker={.},
round-pad=false, exponent-mode = threshold, exponent-thresholds =
-3:5, tight-spacing=true]{siunitx}
\usepackage{subcaption}
\usepackage{bbding}
\usepackage{pifont}

\usepackage{xcolor}
\usepackage{mathtools}

\usepackage{algorithm}
\usepackage[noend]{algorithmic}
\usepackage{makecell}

\input{defs-code}
\newcounter{margincounter}

\newcommand{\NA}[1]{#1}

\newcommand{\DefMacro}[2]{%
   \expandafter\newcommand\csname rmk-#1\endcsname{#2}%
}
\newcommand{\UseMacro}[1]{\csname rmk-#1\endcsname}

\newcommand{\Space}[1]{}

\newcommand{\Mar}[1]{{\color{orange}\bfseries [[MDA: #1]]}}

\definecolor{mintbg}{rgb}{0.94, 1.0, 1.0} 

\newcommand{\mypara}[1]{\vspace{.03in}\noindent \textbf{#1.}}

\newcommand{\eg}{e.g.}
\newcommand{\ie}{i.e.}
\newcommand{\etal}{et al.}

\newcommand{\Code}[1]{{\small\ifmmode{\texttt{#1}}\else$\texttt{#1}$\fi}}
\newcommand{\CodeIn}[1]{{\small\ifmmode{\mathtt{#1}}\else$\mathtt{#1}$\fi}}
\newcommand{\ColorBack}[1]{%
  \begingroup \setlength{\fboxsep}{0pt}
}

\definecolor{gray}{RGB}{211,211,211}

\newcommand{\Contrib}[1]{$\star$~#1}

\newcommand{\Cmodel}{Abstract Input\xspace}
\newcommand{\CMODEL}{AbstractInputs\xspace}

\newcommand{\cmodels}{abstract inputs\xspace}
\newcommand{\cmodel}{abstract input\xspace}
\newcommand{\invariant}{constraint\xspace}
\newcommand{\invariants}{\invariant{}s\xspace}
\newcommand{\Invariant}{Constraint\xspace}
\newcommand{\Invariants}{\Invariant{}s\xspace}

\newcommand{\RQ}[1]{RQ#1\xspace}
\newcommand{\RQs}{RQs\xspace}

\newcommand{\RepoURL}{\url{https://github.com/ncsu-swat/centaur}}

\let\oldtexttt\texttt
\renewcommand{\texttt}[1]{{\small\oldtexttt{#1}}}

\newenvironment{packed_itemize}{
\vspace{-0.7ex}
\begin{list}{\labelitemi}{\leftmargin=0.1em}
 \setlength{\itemsep}{-1em}
 \setlength{\parskip}{0pt}
 \setlength{\parsep}{0pt}
 \setlength{\headsep}{0pt}
 \setlength{\topskip}{0pt}
 \setlength{\topmargin}{0pt}
 \setlength{\topsep}{0pt}
 \setlength{\partopsep}{0pt}
 }{\end{list}
\vspace{-0.7ex}
}

\definecolor{bhisRowColor}{HTML}{64b5cd}

\newcolumntype{C}[1]{>{\centering\arraybackslash}p{#1}}

\newcommand{\nAPIsSL}{\deleted{455}\added{512}}

\newcommand{\nAPIsTF}{\deleted{329}\added{350}} 
\newcommand{\cohensDvalueTF}{\deleted{0.09}\added{0.14}} 

\newcommand{\nAPIsAC}{\deleted{118}\added{124}} 

\newcommand{\nAPIsPF}{\deleted{169}\added{192}} 
\newcommand{\tTestPvaluePF}{\deleted{1.84e-306}\added{0}} 
\newcommand{\cohensDvaluePF}{\deleted{16.94}\added{17.05}} 

\newcommand{\nAPIsSLtf}{\deleted{408}\added{522}}

\newcommand{\nAPIsTFtf}{\deleted{292}\added{333}} 
\newcommand{\cohensDvalueTFtf}{\deleted{0.17}\added{0.16}} 

\newcommand{\nAPIsACtf}{\deleted{87}\added{145}} 

\newcommand{\nAPIsPFtf}{\deleted{94}\added{157}} 
\newcommand{\tTestPvaluePFtf}{\deleted{1.57e-77}\added{6e-128}} 
\newcommand{\cohensDvaluePFtf}{\deleted{4.77}\added{4.76}} 

\newcommand{\tTestPvalueTFVal}{4.12e-153} 
\newcommand{\cohensDvalueTFVal}{3.90} 

\newcommand{\tTestPvalueACVal}{4.91e-37} 
\newcommand{\cohensDvalueACVal}{1.81} 

\newcommand{\tTestPvaluePFVal}{2.24e-96} 
\newcommand{\cohensDvaluePFVal}{2.43} 

\newcommand{\tTestPvalueTFtfVal}{0} 
\newcommand{\cohensDvalueTFtfVal}{5.11} 

\newcommand{\tTestPvalueACtfVal}{2.02e-06} 
\newcommand{\cohensDvalueACtfVal}{0.53} 

\newcommand{\tTestPvaluePFtfVal}{7.22e-07} 
\newcommand{\cohensDvaluePFtfVal}{0.55} 

\newcommand{\dl}{DL\xspace}

\newcommand{\torch}{\textsc{Pytorch}\xspace}
\newcommand{\tf}{\textsc{TensorFlow}\xspace}

\newcommand{\torchVersion}{2.7.1}
\newcommand{\tfVersion}{2.19.0}
\newcommand{\docter}{\textsc{DocTer}\xspace}
\newcommand{\acetest}{\textsc{ACETest}\xspace}
\newcommand{\freefuzz}{\textsc{FreeFuzz}\xspace}
\newcommand{\titanfuzz}{\textsc{TitanFuzz}\xspace}
\newcommand{\deeprel}{\textsc{DeepREL}\xspace}

\newcommand{\pathfinder}{\textsc{Pathfinder}\xspace}
\newcommand{\tname}{\textsc{Centaur}\xspace}

\newcommand{\sota}{SoTA\xspace} 

\newcommand{\torchlinalgsolveexE}{6}
\newcommand{\torchlinalgsolveexVR}{0.0\%}
\newcommand{\torchlinalgsolveexMG}{0.64}
\newcommand{\torchlinalgsolveexMS}{0.70}

\newcommand{\torchfloorE}{4}
\newcommand{\torchfloorVR}{4.1\%}
\newcommand{\torchfloorMG}{0.18}
\newcommand{\torchfloorMS}{0.02}

\newcommand{\torchnnSoftplusE}{2}
\newcommand{\torchnnSoftplusVR}{23.0\%}
\newcommand{\torchnnSoftplusMG}{0.09}
\newcommand{\torchnnSoftplusMS}{0.02}

\newcommand{\torchnninitsparseE}{5}
\newcommand{\torchnninitsparseVR}{2.0\%}
\newcommand{\torchnninitsparseMG}{0.04}
\newcommand{\torchnninitsparseMS}{0.50}

\newcommand{\torchnniniteyeE}{2}
\newcommand{\torchnniniteyeVR}{17.0\%}
\newcommand{\torchnniniteyeMG}{0.01}
\newcommand{\torchnniniteyeMS}{0.01}

\newcommand{\torchlcmE}{5}
\newcommand{\torchlcmVR}{0.0\%}
\newcommand{\torchlcmMG}{0.07}
\newcommand{\torchlcmMS}{0.04}

\newcommand{\torchviewasrealE}{2}
\newcommand{\torchviewasrealVR}{24.0\%}
\newcommand{\torchviewasrealMG}{0.01}
\newcommand{\torchviewasrealMS}{0.01}

\newcommand{\torchnarrowE}{4}
\newcommand{\torchnarrowVR}{0.0\%}
\newcommand{\torchnarrowMG}{0.75}
\newcommand{\torchnarrowMS}{0.05}

\newcommand{\torchabsE}{5}
\newcommand{\torchabsVR}{11.3\%}
\newcommand{\torchabsMG}{0.22}
\newcommand{\torchabsMS}{0.08}

\newcommand{\torchchannelshuffleE}{4}
\newcommand{\torchchannelshuffleVR}{2.0\%}
\newcommand{\torchchannelshuffleMG}{0.08}
\newcommand{\torchchannelshuffleMS}{0.02}

\newcommand{\tensorflownncreluE}{3}
\newcommand{\tensorflownncreluVR}{0.0\%}
\newcommand{\tensorflownncreluMG}{0.03}
\newcommand{\tensorflownncreluMS}{0.02}

\newcommand{\tensorflowsparseeyeE}{2}
\newcommand{\tensorflowsparseeyeVR}{35.4\%}
\newcommand{\tensorflowsparseeyeMG}{0.01}
\newcommand{\tensorflowsparseeyeMS}{0.01}

\newcommand{\tensorflowexperimentalnumpyanyE}{2}
\newcommand{\tensorflowexperimentalnumpyanyVR}{0.0\%}
\newcommand{\tensorflowexperimentalnumpyanyMG}{0.03}
\newcommand{\tensorflowexperimentalnumpyanyMS}{0.02}

\newcommand{\tensorfloweyeE}{2}
\newcommand{\tensorfloweyeVR}{0.0\%}
\newcommand{\tensorfloweyeMG}{0.05}
\newcommand{\tensorfloweyeMS}{0.61}

\newcommand{\tensorflownnisotonicregressionE}{2}
\newcommand{\tensorflownnisotonicregressionVR}{2.0\%}
\newcommand{\tensorflownnisotonicregressionMG}{0.05}
\newcommand{\tensorflownnisotonicregressionMS}{0.02}

\newcommand{\tensorflowimagessimE}{3}
\newcommand{\tensorflowimagessimVR}{0.0\%}
\newcommand{\tensorflowimagessimMG}{0.10}
\newcommand{\tensorflowimagessimMS}{0.22}

\newcommand{\tensorflowmathinvertpermutationE}{2}
\newcommand{\tensorflowmathinvertpermutationVR}{0.0\%}
\newcommand{\tensorflowmathinvertpermutationMG}{0.18}
\newcommand{\tensorflowmathinvertpermutationMS}{0.05}

\newcommand{\tensorflowlinalginvE}{3}
\newcommand{\tensorflowlinalginvVR}{0.0\%}
\newcommand{\tensorflowlinalginvMG}{0.02}
\newcommand{\tensorflowlinalginvMS}{0.02}

\newcommand{\tensorflowsignalfftE}{5}
\newcommand{\tensorflowsignalfftVR}{24.0\%}
\newcommand{\tensorflowsignalfftMG}{0.02}
\newcommand{\tensorflowsignalfftMS}{0.02}

\newcommand{\tensorflowuniquewithcountsE}{2}
\newcommand{\tensorflowuniquewithcountsVR}{10.0\%}
\newcommand{\tensorflowuniquewithcountsMG}{0.16}
\newcommand{\tensorflowuniquewithcountsMS}{0.14}

\newcommand{\totalLeftIGNum}{31}

\newcommand{\totalLeftIGPct}{88.6\%}

\newcommand{\totalLeftISDen}{80}
\newcommand{\totalLeftISPct}{98.8\%}
\newcommand{\totalRightIGNum}{32}

\newcommand{\totalRightIGPct}{100.0\%}

\newcommand{\totalRightISDen}{88}
\newcommand{\totalRightISPct}{89.8\%}

\newcommand{\numDaggeredRecall}{27}
\newcommand{\numDaggeredPrecision}{22}

\newcommand{\torchBugs}{13}
\newcommand{\tfBugs}{13}

\newcommand{\totalBugs}{26}

\newcommand{\torchBugsSubmitted}{13}
\newcommand{\tfBugsSubmitted}{13}
\newcommand{\totalBugsSubmitted}{26}

\newcommand{\torchBugsConfirmed}{8}
\newcommand{\tfBugsConfirmed}{10}
\newcommand{\totalBugsConfirmed}{18}

\newcommand{\torchBugsRejected}{3}
\newcommand{\tfBugsRejected}{1}
\newcommand{\totalBugsRejected}{4}

\newcommand{\torchBugsPreviouslyFixed}{2}
\newcommand{\tfBugsPreviouslyFixed}{0}
\newcommand{\totalBugsPreviouslyFixed}{2}

\newcommand{\torchBugsPending}{2}
\newcommand{\tfBugsPending}{2}
\newcommand{\totalBugsPending}{4}

\newcommand{\torchBugsFixed}{1}
\newcommand{\tfBugsFixed}{0}
\newcommand{\totalBugsFixed}{1}

\newcommand{\torchBugsCrash}{1}
\newcommand{\tfBugsCrash}{6}
\newcommand{\totalBugsCrash}{7}

\newcommand{\torchBugsInconsistent}{9}
\newcommand{\tfBugsInconsistent}{6}
\newcommand{\totalBugsInconsistent}{15}

\newcommand{\torchBugsOverflow}{2}
\newcommand{\tfBugsOverflow}{1}
\newcommand{\totalBugsOverflow}{3}

\newcommand{\torchBugsNan}{1}
\newcommand{\tfBugsNan}{0}
\newcommand{\totalBugsNan}{1}

\newcommand{\moreCovThanTitanfuzz}{\deleted{236}\added{203}} 
\newcommand{\moreCovThanACETest}{\deleted{42}\added{150}} 
\newcommand{\moreCovThanPathfinder}{\deleted{9492}\added{9,608}} 

\newcommand{\torchnnReLUE}{3}
\newcommand{\torchnnReLUVR}{66.0\%}
\newcommand{\torchnnReLUMG}{0.04}
\newcommand{\torchnnReLUMS}{0.04}

\newcommand{\torchargmaxE}{4}
\newcommand{\torchargmaxVR}{32.0\%}
\newcommand{\torchargmaxMG}{0.05}
\newcommand{\torchargmaxMS}{0.03}

\newcommand{\torchnnfunctionalalphadropoutE}{3}
\newcommand{\torchnnfunctionalalphadropoutVR}{0.0\%}
\newcommand{\torchnnfunctionalalphadropoutMG}{0.03}
\newcommand{\torchnnfunctionalalphadropoutMS}{0.03}

\newcommand{\tensorflowmathsigmoidE}{2}
\newcommand{\tensorflowmathsigmoidVR}{48.0\%}
\newcommand{\tensorflowmathsigmoidMG}{0.02}
\newcommand{\tensorflowmathsigmoidMS}{0.03}

\newcommand{\torchlinalghouseholderproductE}{5}
\newcommand{\torchlinalghouseholderproductVR}{0.0\%}
\newcommand{\torchlinalghouseholderproductMG}{0.18}
\newcommand{\torchlinalghouseholderproductMS}{0.09}

\newcommand{\torchspecialexpitE}{2}
\newcommand{\torchspecialexpitVR}{42.4\%}
\newcommand{\torchspecialexpitMG}{0.15}
\newcommand{\torchspecialexpitMS}{0.15}

\newcommand{\tensorflowlinalgmatrixrankE}{2}
\newcommand{\tensorflowlinalgmatrixrankVR}{13.0\%}
\newcommand{\tensorflowlinalgmatrixrankMG}{0.38}
\newcommand{\tensorflowlinalgmatrixrankMS}{0.38}

\newcommand{\tensorflowmaketensorprotoE}{3}
\newcommand{\tensorflowmaketensorprotoVR}{6.0\%}
\newcommand{\tensorflowmaketensorprotoMG}{2.02}
\newcommand{\tensorflowmaketensorprotoMS}{0.21}

\newcommand{\tensorflowuniqueE}{2}
\newcommand{\tensorflowuniqueVR}{10.0\%}
\newcommand{\tensorflowuniqueMG}{0.06}
\newcommand{\tensorflowuniqueMS}{0.01}

\newcommand{\tensorflowimagetransposeE}{4}
\newcommand{\tensorflowimagetransposeVR}{46.0\%}
\newcommand{\tensorflowimagetransposeMG}{0.03}
\newcommand{\tensorflowimagetransposeMS}{0.04}

\newcommand{\validTitanfuzzPtBase}{123k}
\newcommand{\validTitanfuzzPtSlate}{48,320k}
\newcommand{\validTitanfuzzTfBase}{34k}
\newcommand{\validTitanfuzzTfSlate}{56,248k}
\newcommand{\deltaTitanfuzzPt}{\textcolor{blue}{$\uparrow$ 48,197k}}
\newcommand{\deltaTitanfuzzTf}{\textcolor{blue}{$\uparrow$ 56,214k}}
\newcommand{\validAcetestPtBase}{2,092k}
\newcommand{\validAcetestPtSlate}{34,822k}
\newcommand{\validAcetestTfBase}{10,556k}
\newcommand{\validAcetestTfSlate}{14,061k}
\newcommand{\deltaAcetestPt}{\textcolor{blue}{$\uparrow$ 32,730k}}
\newcommand{\deltaAcetestTf}{\textcolor{blue}{$\uparrow$ 3,505k}}
\newcommand{\validPathfinderPtBase}{5,658k}
\newcommand{\validPathfinderPtSlate}{65,479k}
\newcommand{\validPathfinderTfBase}{105k}
\newcommand{\validPathfinderTfSlate}{13,367k}
\newcommand{\deltaPathfinderPt}{\textcolor{blue}{$\uparrow$ 59,820k}}
\newcommand{\deltaPathfinderTf}{\textcolor{blue}{$\uparrow$ 13,262k}}

%% file: defs-code.tex
\definecolor{gray}{RGB}{211,211,211}
\newcommand{\jbasicstyle}{\small\sffamily} 

\newcommand{\jnumberstyle}{\scriptsize}

\definecolor{mymauve}{rgb}{0.58,0,0.82}

\lstdefinelanguage{pseudo}
{
  morekeywords={},
  keywordstyle=\bfseries,
  lineskip=-0.1em,
  numbers=left, 
  numberstyle=\jnumberstyle,
  numbersep=4pt,
  basicstyle=\jbasicstyle,
  breaklines=true,
  breakautoindent=true,
  tabsize=2,
  columns=fullflexible,
  morecomment=*[l][\textsl]{//},
  mathescape=true,
  xleftmargin=10pt,
}

\lstdefinelanguage{todo-comment}
{
  morekeywords={},
  keywordstyle=\bfseries,
  lineskip=-0.1em,
  numbers=none,
  basicstyle=\jbasicstyle,
  breaklines=true,
  breakautoindent=true,
  tabsize=2,
  columns=fullflexible,
  morecomment=*[l][\textsl]{//},
  mathescape=true,
  xleftmargin=-10pt,
}

\lstdefinelanguage{java-pretty}
{
  language=java,
  numbers=none,                 
  basicstyle=\scriptsize\ttfamily,
  numberstyle=\scriptsize\tiny,
  numbers=left,
  breaklines=true,
  columns=fullflexible,
  xleftmargin=3pt,
  showstringspaces=false,
  numbersep=1.4pt,
}

\makeatletter
\newenvironment{btHighlight}[1][]
{\begingroup\tikzset{bt@Highlight@par/.style={#1}}\begin{lrbox}{\@tempboxa}}
{\end{lrbox}\bt@HL@box[bt@Highlight@par]{\@tempboxa}\endgroup}
\newcommand\btHL[1][]{%
  \begin{btHighlight}[#1]\bgroup\aftergroup\bt@HL@endenv%
}
\def\bt@HL@endenv{%
  \end{btHighlight}%
  \egroup
}
\newcommand{\bt@HL@box}[2][]{%
  \tikz[#1]{%
    \pgfpathrectangle{\pgfpoint{1pt}{0pt}}{\pgfpoint{\wd #2}{\ht #2}}%
    \pgfusepath{use as bounding box}%
    \node[anchor=base west, fill=orange!30,outer sep=0pt,inner xsep=1pt, inner ysep=0pt, rounded corners=1pt, minimum height=\ht\strutbox,#1]{\raisebox{1pt}{\strut}\strut\usebox{#2}};
  }%
}
\newenvironment{btHighlightLine}[1][]
{\begingroup\tikzset{bt@HighlightLine@par/.style={#1}}\begin{lrbox}{\@tempboxa}}
{\end{lrbox}\bt@HLLine@box[bt@HighlightLine@par]{\@tempboxa}\endgroup}
\newcommand\btHLLine[1][]{%
  \begin{btHighlightLine}[#1]\bgroup\aftergroup\bt@HLLine@endenv%
}
\def\bt@HLLine@endenv{%
  \end{btHighlightLine}%
  \egroup
}
\newcommand{\bt@HLLine@box}[2][]{%
  \tikz[#1]{%
    \pgfpathrectangle{\pgfpoint{0pt}{-1pt}}{\pgfpoint{\wd #2}{\ht #2}}%
    \pgfusepath{use as bounding box}%
    \node[anchor=base west, fill=orange!30,outer sep=0pt,inner xsep=0pt, inner ysep=0pt, minimum height=\ht\strutbox+3pt, minimum width=\linewidth,#1] (line-bg) {};
    \node[right = 0 of line-bg.west, outer sep=0pt, inner xsep=0pt, inner ysep=0pt]{\raisebox{0pt}{\strut}\strut\usebox{#2}};
  }%
}
\makeatother
\lstdefinelanguage{java-diff}
{
  language=java,
  columns=fullflexible,
  xleftmargin=3pt,
  numbers=left,
  numbersep=1.4pt,
  basicstyle=\scriptsize\ttfamily,
  numberstyle=\scriptsize\tiny\color{darkgray},
  breaklines=true,
  showstringspaces=false,
  moredelim=**[il][{\btHLLine[fill=red!25]}]{(-L)},
  moredelim=**[il][{\btHLLine[fill=green!25]}]{(+L)},
  moredelim=**[is][{\btHL[fill=red!20]}]{(-W<)}{(>W-)},
  moredelim=**[is][{\btHL[fill=green!20]}]{(+W<)}{(>W+)},
  literate=
         {-}{-}{1}
}

\lstdefinelanguage{AspectJ}[]{java-pretty}{
    morekeywords={declare, pointcut, aspect, before, around, after, returning, throwing, call, execution, this, target, args, within, withincode, get, set, initialization, preinitialization, staticinitialization, handler, adviceexecution, cflow, cflowbelow, if, proceed, event, ere, ltl, @match, creation, @violation},
    moredelim=[is][\textcolor{darkgray}]{\%\%}{\%\%},
    moredelim=[il][\textcolor{darkgray}]{§§},
    basicstyle=\scriptsize\ttfamily,
    numbersep=1pt,
    numbers=left,
    numberstyle=\scriptsize\tiny\color{darkgray},    
    breaklines=true,
    escapeinside={(*@}{@*)},
    showstringspaces=false,
    xleftmargin=3pt
}

%% file: intro.tex
\section{Introduction}
\label{sec:intro}
AI-enabled applications are prolific today.  Deep Learning (\dl) libraries, such
as \tf~\cite{abadi2016tensorflow}, and \torch~\cite{paszke2019pytorch} provide
the learning components to build these applications. \dl libraries are very
complex systems and prone to bugs.
Finding bugs in \dl\ libraries is an important and challenging problem that
recent prior work has
tackled~\cite{Jia_ETAL_JSS21,Xie_ETAL_ISSTA22,Wei2022,Chen_ETAL_TOSEM23}.

Despite the impressive advances of recent testing techniques, there still is
much room to improve their ability to find \emph{deep bugs}, i.e., bugs
associated with the core functionality of the APIs. DL library APIs often do not
come with formal input specifications.
Hence, most existing techniques are either oblivious to the input constraints of
the APIs under test (\eg{}, \deeprel~\cite{Deng_ETAL_FSE22} and
\freefuzz~\cite{Wei2022}) or are imprecise in modeling input constraints (\eg{},
\docter~\cite{Xie_ETAL_ISSTA22} and \acetest~\cite{Shi_ETAL_ISSTA23}). Prior
techniques for mining API input constraints use either API documentation (\eg{},
\docter) or symbolic code analysis (e.g., \acetest\ and
\pathfinder~\cite{Kim2025LightweightConcolic}). Mining constraints from either
source can be challenging. API documentation is often incomplete and sometimes
specifications can be scattered across multiple documents. On the other hand,
symbolic techniques (e.g., by analyzing path conditions) can be costly and
imprecise.
%
Due to these challenges\Space{ of these challenges}, such techniques often mostly reveal shallow
bugs, \ie{}, bugs associated with input validity checkers of the APIs as opposed
to bugs related to the deeper functionality of the API. To reveal deep
(semantic) bugs, it is important to accurately model the input constraints
for each API in \dl\ library.

{\let\texttt\oldtexttt
\begin{figure*}[t]
  \centering
  \footnotesize
  \begin{subfigure}[t]{0.48\textwidth}
    \vspace{0pt}
    \input{figures/api_doc_subfig.tex}
    \vspace{-2pt}
    \caption{API documentation (\texttt{torch.add})}
    \label{fig:doc_example:api_doc}
  \end{subfigure}
  \hfill
  \begin{subfigure}[t]{0.5\textwidth}
    \vspace{0pt}
    \input{figures/broadcasting_doc.tex}
    \vspace{-2pt}    
    \caption{Broadcasting semantics (\texttt{torch.add})}
    \label{fig:doc_example:broadcasting_semantics}
  \end{subfigure}
  \vspace{-2ex}
  \caption{Documentation for PyTorch API and broadcasting semantics.}
  \vspace{-3ex}
\end{figure*}
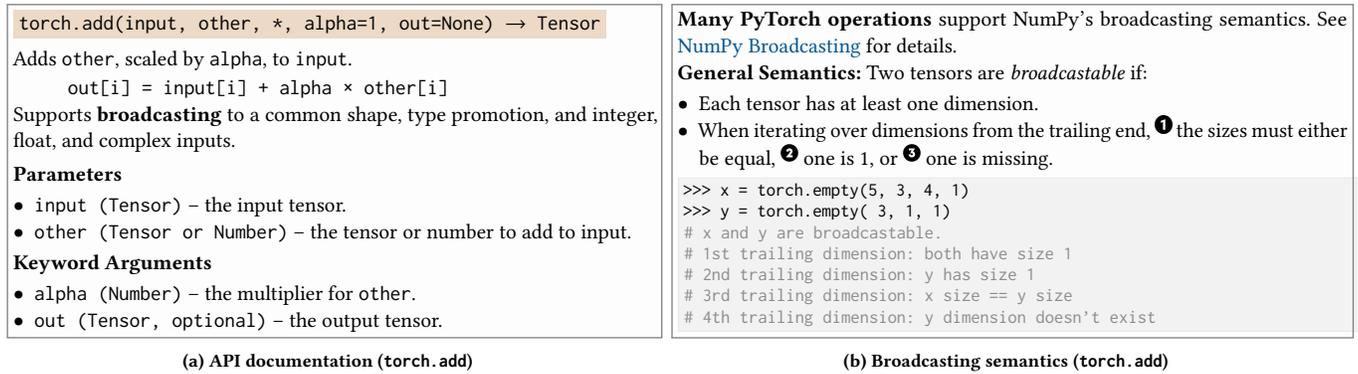
}



%
\mypara{Our Work}
We propose \tname, the first \emph{neurosymbolic} technique to test \dl\ APIs
using dynamically-learned input \invariants. 
%
\tname takes as input (1) a DL API, (2) a set of valid inputs for
that API and (3) a budget on the number of inputs to generate, and
reports likely bugs as output.
The approach works in three steps. 
First, \tname uses a grammar characterizing \dl\ library API constraints related to
input parameters.  A \emph{rule} is a candidate of a constraint for a
combination of API parameters, \eg{}, the last dimensions of two
tensor parameters should match. 
Second, \tname\ learns input \emph{constraints} for an API by mining valid inputs,
identifying which of the candidate rules obtained in the previous phase hold.
Third, \tname\ uses an SMT solver (\eg{}, Z3~\cite{demoura2008z3}) to generate
models for abstract API inputs 
and a sampling strategy to generate concrete values from the abstract inputs.

\tname\ needs to overcome several key challenges to be effective. First, the
space of candidate rules is prohibitively large, which makes simple enumeration
and validation of rules impractical. Second, it is difficult to obtain a
sizeable number of inputs for each API (e.g., PyTorch has several hundreds of
APIs) for learning input constraints. Ideally we need multiple and diverse valid
inputs to obtain high quality constraints. Third, the constraints learned from a
sample of inputs may not be generalizable to the entire API, and include
spurious or incorrect rules. Fourth, typically SMT solvers only output one
concrete solution for a given set of constraints, which limits the diversity of
inputs and is also time-consuming if we want to generate multiple inputs.

\tname addresses these challenges as follows. First, we design a novel
grammar that represents first-order logic formulae and tensor-related
properties (such as shape and data type). Then, we use a Large
Language Model (LLM) to take this grammar specification and API
documentation as inputs and generate candidate rules. Our intuition
is that LLMs embody DL API-specific domain knowledge and can
generate more relevant properties compared with a simple enumerator
like grammarinator~\cite{hodovan2018grammarinator}. Further, by using
a grammar, we can constrain the LLM outputs to syntactically valid
rules. Second, we use LLMs to generate a small set of valid
inputs for each API for constraint learning, and use custom mutators
to diversify the inputs.
Third, we develop a refinement strategy to systematically eliminate
spurious or redundant rules. \NA{Finally, we show how we can use SMT solvers to
generate abstract inputs that satisfy the learned constraints. The abstract
inputs represent valid input sub-spaces that can be efficiently sampled
from. We also develop optimizations to diversify the solver-generated abstract 
inputs.} 


\mypara{Results}
\enlargethispage{\baselineskip}
We conduct three experiments to evaluate \tname. First, we evaluate the
constraints that \tname\ generates quantitatively and qualitatively. We find
that \tname\ generates\Space{ precise constraints for the APIs that we analyze and
generates} more accurate and more precise constraints than prior approaches.
Second, we show that \tname\ outperforms \acetest~\cite{Shi2023ACETest},
\pathfinder~\cite{Kim2025LightweightConcolic}, and
\titanfuzz~\cite{Deng_ETAL_ISSTA23} on both branch coverage and validity ratio
(\ie, \NA{ratio of generated inputs that satisfy API-specific validity
constraints}) -- two established metrics adopted in the literature to compare
\dl\ API-level fuzzers.
%
For example, \tname\ achieves a validity ratio of 98.2\% for \torch
and of 99.24\% for \tf; these numbers are at least 28.2 and at most 
77.7 percentage points higher than the validity ratio of 
the other comparison baselines. Third, \tname\ finds \totalBugs\ bugs
in \torch\ and \tf. Of these, \totalBugsConfirmed\ bugs have been confirmed by
developers, and one has been fixed at the time of writing this paper.

\mypara{Contributions} We make the following contributions:
\begin{itemize}[topsep=0ex,itemsep=0pt,leftmargin=1.3em]
\item[\Contrib{}] \textbf{Idea.} We propose \tname, the first neurosymbolic
  API-level fuzzer for \dl\ libraries \NA{that combines grammar-driven LLM-assisted
  constraint learning (the neural part) with SMT-based test input
  generation (the symbolic part)};
\item[\Contrib{}] \textbf{Evaluation.} We comprehensively evaluate \tname\ using
  standard metrics from the literature and compare \tname\ against \sota\
  techniques (\eg, \acetest, \pathfinder, and \titanfuzz). Results indicate the
  superiority of \tname\ and its ability to reveal deep bugs in \dl\ libraries;
\item[\Contrib{}] \textbf{Bugs.}  \NA{Using \tname, we found \totalBugs\
  previously unknown bugs in \torch\ and \tf, \totalBugsConfirmed\ of which have
  been confirmed by developers};
\item[\Contrib{}] \textbf{Tool.} \tname\ is publicly-available at
	\RepoURL. 
\end{itemize}

%% file: figures/api_doc_subfig.tex
\begingroup
\setlength{\fboxsep}{2pt}
\setlength{\fboxrule}{0.5pt}
\fcolorbox{black!50}{gray!5}{%
\begin{minipage}{\linewidth}%
  \small
  \colorbox{brown!30}{\texttt{torch.add(input, other, *, alpha=1, out=None) → Tensor}}

  \vspace{1ex}
  Adds \texttt{other}, scaled by \texttt{alpha}, to \texttt{input}.

  \vspace{0.3ex}
  \hspace{6ex}\texttt{out[i] = input[i] + alpha × other[i]}

  Supports \textbf{broadcasting} to a common shape, type promotion, and integer, float, and complex inputs.

  \vspace{0.8ex}
  \textbf{Parameters}

  \begin{itemize}[leftmargin=*]
    \item \texttt{input (Tensor)} – the input tensor.
    \item \texttt{other (Tensor or Number)} – the tensor or number to add to input.
  \end{itemize}

  \textbf{Keyword Arguments}

  \begin{itemize}[leftmargin=*]
    \item \texttt{alpha (Number)} – the multiplier for \texttt{other}.
    \item \texttt{out (Tensor, optional)} – the output tensor.
  \end{itemize}
\end{minipage}%
}
\endgroup

%% file: figures/broadcasting_doc.tex
{\small
\begingroup
\setlength{\fboxsep}{2pt}
\setlength{\fboxrule}{0.5pt}
\fcolorbox{black!50}{gray!5}{%
\begin{minipage}{\linewidth}%
  \textbf{Many PyTorch operations} support NumPy’s broadcasting semantics.
  See \href{https://numpy.org/doc/stable/user/basics.broadcasting.html}{NumPy Broadcasting} for details.

  \vspace{0.1ex}
  \textbf{General Semantics:} Two tensors are \emph{broadcastable} if:

  \begin{itemize}[leftmargin=*]
    \item Each tensor has at least one dimension.
    \item When iterating over dimensions from the trailing end, 
        \raisebox{0.4ex}{\includegraphics[height=1.9ex]{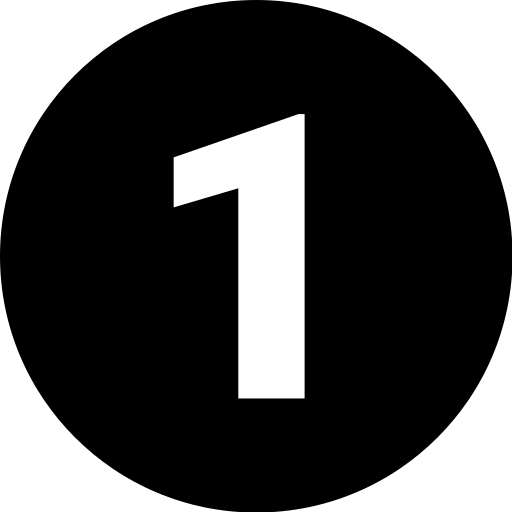}}~the sizes must either be equal, \raisebox{0.4ex}{\includegraphics[height=1.9ex]{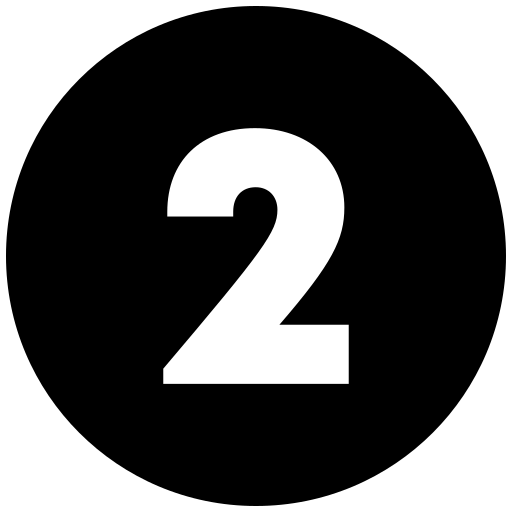}}~one is 1, or \raisebox{0.4ex}{\includegraphics[height=1.9ex]{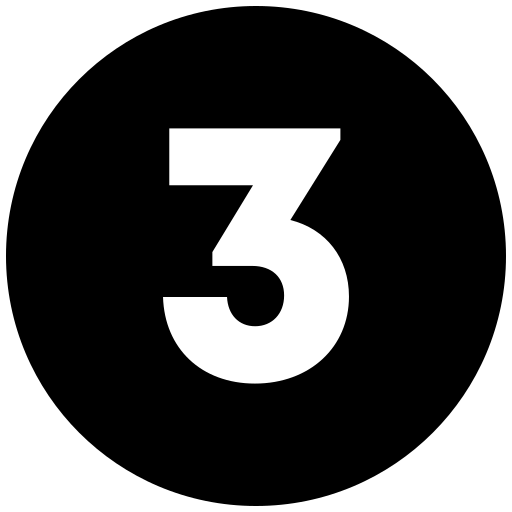}}~one is missing.
  \end{itemize}

\begingroup
\setlength{\fboxsep}{2pt}
\setlength{\fboxrule}{0.3pt}
\fcolorbox{gray!50}{gray!30}{%
\begin{minipage}{\linewidth}%
  \footnotesize\ttfamily
  \textgreater{}\textgreater{}\textgreater{} x = torch.empty(5, 3, 4, 1)\\
  \textgreater{}\textgreater{}\textgreater{} y = torch.empty(   3, 1, 1)\\
  \textcolor{darkgray!60}{\# x and y are broadcastable.}\\
  \textcolor{darkgray!60}{\# 1st trailing dimension: both have size 1}\\
  \textcolor{darkgray!60}{\# 2nd trailing dimension: y has size 1}\\
  \textcolor{darkgray!60}{\# 3rd trailing dimension: x size == y size}\\
  \textcolor{darkgray!60}{\# 4th trailing dimension: y dimension doesn't exist}
\end{minipage}%
}%
\endgroup

\end{minipage}%
}%
\endgroup
}

%% file: example.tex
\begin{figure*}[!htb]
  \centering
  \includegraphics[width=0.92\linewidth]{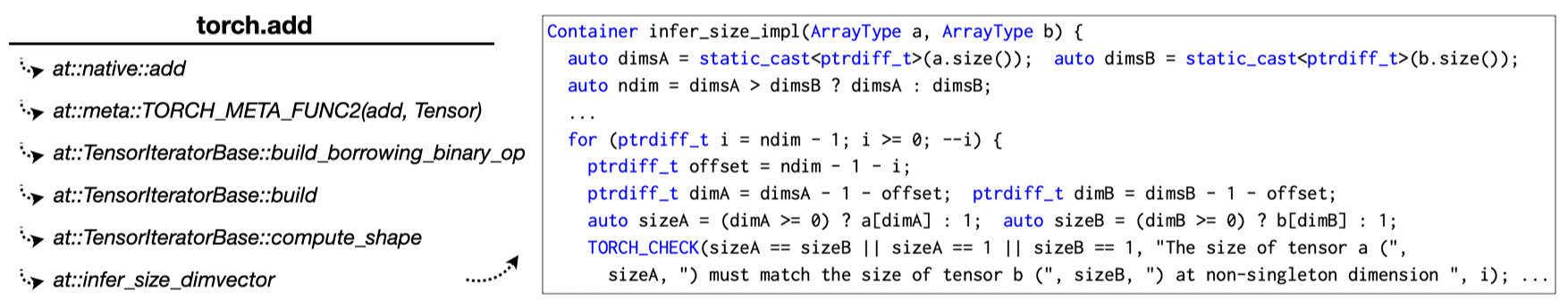}
  \vspace{-2ex}
  \caption{Native code for \texttt{\footnotesize torch.add} API.}
  \label{fig:native_code}
  \vspace{-2.5ex}  
\end{figure*}

{\let\texttt\oldtexttt
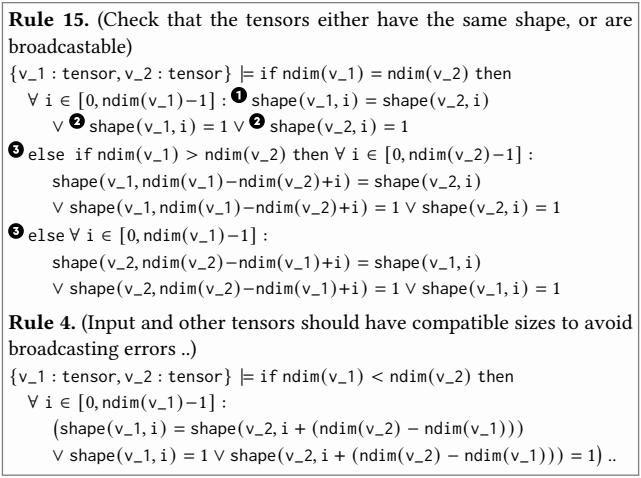
\begin{figure}[t]
\centering
\vspace{0pt}
\input{figures/formal_rule_subfig.tex}
\vspace{-4mm}
\caption{Formal rules \tname learned for broadcasting semantics.}
\label{fig:rule_code:formal_rules}
\vspace{-2.5ex}
\end{figure}
}


\section{Motivating Example}
\label{sec:example}
We present a motivating example of how \tname can be used to generate valid
inputs for PyTorch's \texttt{torch.add} API. Figure~\ref{fig:doc_example:api_doc}
presents the documentation for PyTorch's \texttt{torch.add} API.\Space{ It
defines the function signature, parameters, and keyword (optional) arguments.}
The API takes two tensors as input--\texttt{input} and \texttt{other}--and adds
them. The \texttt{other} tensor can be multiplied by \texttt{alpha}, if that
parameter is provided. The resulting tensor can be optionally returned as
\texttt{out} variable. The two input tensors should satisfy the \emph{broadcasting}
semantics~\cite{pytorch-broadcasting}, which is described by the documentation
that Figure~\ref{fig:doc_example:broadcasting_semantics} shows. Intuitively, we
say that two tensors are broadcastable if, starting from the trailing dimension,
one of the three conditions
is met for all dimensions (see \ding{182}, \ding{183}, and \ding{184}). For
example, tensors with shape \CodeIn{(5, 3, 4, 1)} and \CodeIn{(3, 1, 1)} are
broadcastable as explained in comments. 
%
Inputs that do not satisfy this property lead to a runtime error, and
notably, this semantics applies to \emph{many} APIs
(e.g., \texttt{torch.mul}, \texttt{torch.sub}, and \texttt{torch.div}).
For example, if \CodeIn{torch.randn(3, 4)} and \CodeIn{torch.randn(2,
4)} are passed to \CodeIn{torch.mul}, it raises an error ``size of tensor
\CodeIn{a} (3) must match the size of tensor \CodeIn{b} (2) at non-singleton
dimension 0''.


Testing this API is challenging for two reasons. \underline{First}, the
description of properties can be \emph{scattered} throughout the documentation.
Note that the API documentation briefly mentions broadcasting semantics; the
definition of the actual property appears in a separate documentation. More
often, the property may not be explicitly described in any documentation, i.e.,
properties can be \emph{implicit}. For example,
\texttt{torch.nn.ConvTranspose2d} has an implicit constraint that
\texttt{output\_padding} should be less than \texttt{stride}, which is not
described in documentation. \underline{Second}, properties are described in
natural/informal language and often include multiple parameters of an API.


In prior work, Xie \etal{}~\cite{Xie_ETAL_ISSTA22} extracted constraints from
documentation of DL APIs. However, they extracted constraints only from
parameter descriptions in API documentation
\added{; their approach} cannot extract
properties that are scattered in the documentation or are implicit Also, their
constraint extraction is based on syntactic patterns and each constraint only involves a
\emph{single} parameter at a time.\Space{While they can express types and some properties
of a parameter, they cannot express broadcasting semantics.} For example, the
following shows their constraints for \texttt{torch.add} API: 

\begin{center}
  \renewcommand{\arraystretch}{0.95} 
  \setlength{\tabcolsep}{1pt}
  \hspace{1mm}
  \begin{tabular}{@{}p{0.48\linewidth}@{\hspace{0.01\linewidth}}p{0.45\linewidth}@{}}
    \textbf{\texttt{input}}: \textit{tensor\_t}=\texttt{torch.tensor}
    &
    \textbf{\texttt{other}}: \textit{tensor\_t}=\texttt{torch.tensor}
    \\
    \textbf{\texttt{alpha}}: \textit{dtype}=int, \textit{ndim}=0
    &
    \textbf{\texttt{out}}: \textit{tensor\_t}=\texttt{torch.tensor}
  \end{tabular}
\end{center}

\noindent Each italicized keyword corresponds to a constraint: \emph{tensor\_t}
constrains a tensor type, and \emph{dtype} and \emph{ndim} constrain the data
type and number of dimensions of tensor. In total, they have five constraints
for the API. As the example shows, they only capture simple tensor type
constraints that are explicit in parameter descriptions. Also, their constraints
are often imprecise (e.g., wrong \textit{ndim} constraint for the number-typed 
\textbf{\texttt{alpha}} parameter)
as based on syntactic patterns, and cannot express relational
properties over multiple parameters, like broadcasting semantics.


To address the challenges of using only documentation, recent work proposed
source code-based constraint extraction~\cite{Shi_ETAL_ISSTA23,
Kim2025LightweightConcolic}. However, their constraints are not as
\emph{precise}. \acetest~\cite{Shi_ETAL_ISSTA23} identifies input validation
path by traversing backward from error-handling function calls. For each
identified path, \acetest incrementally builds constraints based on pre-defined
rules. However, as Figure~\ref{fig:native_code} shows, error handling code can
often be reached after a long sequence of function calls. Also, in this example,
the \texttt{build} function has multiple function calls internally, and both the
\texttt{compute\_shape} and \texttt{infer\_size\_impl} functions have a loop.
This leads to a potential path explosion, and \acetest fails at identifying any
constraints for \texttt{torch.abs} API.
\pathfinder~\cite{Kim2025LightweightConcolic} uses lightweight concolic testing;
it inductively refines path conditions by input executions. However, for the
same explosion reason, \pathfinder fails to construct precise path
conditions for broadcasting semantics. 

\tname addresses these challenges by leveraging the knowledge of LLMs. Due to
the popularity of DL libraries, LLM has sufficient knowledge of valid API
usages. We leverage it to generate formal constraints which include those from
scattered documentations and implicit properties
(\textsection~\ref{tech:rulegen}). Also, \tname defines a novel grammar
(\textsection~\ref{tech:grammar}) that can express relational properties over an
arbitrary number of parameters.
Using this novel combination, \tname effectively generates a rule for
broadcasting semantics as shown in Rule 15 of Figure~\ref{fig:rule_code:formal_rules}. 
As the example shows, the rule precisely captures all three conditions.
It not only checks for the cases where two tensors have the same number of
dimensions (\ding{182} and \ding{183}),
 but also check for the cases where one tensor has more number of
 dimensions (\ding{184}).
Further, \tname eliminates redundant LLM-generated rules to improve precision.
Rule 4 in Figure~\ref{fig:rule_code:formal_rules} shows another generated
rule that is essentially the same as Rule 15, but in a different representation. 
\tname dynamically checks redundancy based on validity ratio and
filters out such rules~\cite{torchadd}. 

\begin{figure*}[!htb]
  \centering
  \includegraphics[width=0.97\linewidth]{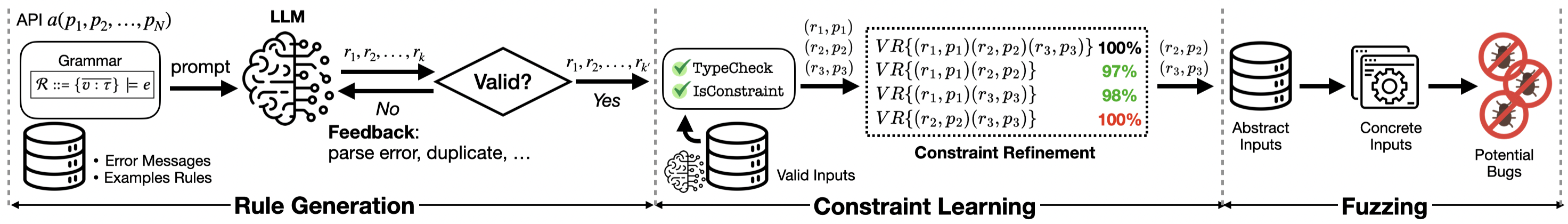}
  \vspace{-1.5ex}
  \caption{\tname's workflow.}
  \label{fig:workflow}
  \vspace{-1ex}  
\end{figure*}

{\let\texttt\oldtexttt
\begin{figure}[t]
 \vspace{-.5ex}  
  \centering

  \begin{subfigure}[t]{\columnwidth}
    \vspace{0pt}
    \input{figures/grammar_subfig.tex}
    \vspace{-1.5ex}
    \caption{Grammar of API constraints}
    \label{fig:grammar}

  \end{subfigure}

  \vspace{.3em}

  \begin{subfigure}[t]{\columnwidth}
    \vspace{0pt}
    \input{figures/typing_rules.tex}
    \vspace{-1.5ex}
    \caption{Typing rules for expressions and constraints}
    \label{fig:typing_rules}
  \end{subfigure}
  \vspace{-2.1ex}  
  \caption{Grammar and typing rules of \tname.}
  \label{fig:grammar_typing}
  \vspace{-3.4ex}
\end{figure}
}

\input{algorithms/rule_generation.tex}

\enlargethispage{\baselineskip}
Using refined rules, \tname automatically converts the grammar-based
definitions to Python code that encodes the rules as SMT formulae.
\tname uses the Python code to dynamically identify constraints (pairs of a rule
and a subset of API parameters that the rule applies to) from the rules. \tname
tests whether the generated rules are satisfied by a set of automatically
generated valid inputs. If so, \tname saves the pair as a
constraint~(\textsection~\ref{tech:refine}). Next, \tname uses the learned
constraints to generate diverse valid inputs by solving the SMT formulae
(\textsection~\ref{tech:model}). With this approach, \tname can generate
valid inputs efficiently (8.9 milliseconds on average per input), and achieves
near 100\% validity ratio with no runtime errors.

%% file: figures/formal_rule_subfig.tex
{\let\texttt\oldtexttt
\small
\noindent
\begingroup
\vspace{1.3ex}
\setlength{\fboxsep}{2pt}
\setlength{\fboxrule}{0.5pt}
\fcolorbox{black!50}{gray!5}{%
\begin{minipage}{\dimexpr\linewidth-2\fboxsep-2\fboxrule\relax}
\vspace{1pt}

\textbf{Rule 15.} (Check that the tensors either have the same shape, or are broadcastable)

{\footnotesize
$\{\texttt{v\_1}: \texttt{tensor}, \texttt{v\_2}: \texttt{tensor}\} \models \texttt{if}\ \texttt{ndim}(\texttt{v\_1}) = \texttt{ndim}(\texttt{v\_2})\ \texttt{then}$\\
\hspace*{1em}$\forall\ \texttt{i} \in [0, \texttt{ndim}(\texttt{v\_1}){-}1] :\raisebox{0.4ex}{\includegraphics[height=1.9ex]{images/number1.png}}~\texttt{shape}(\texttt{v\_1}, \texttt{i}) = \texttt{shape}(\texttt{v\_2}, \texttt{i})$\\
\hspace*{2.35em}$\lor\ \raisebox{0.4ex}{\includegraphics[height=1.9ex]{images/number2.png}}~\texttt{shape}(\texttt{v\_1}, \texttt{i}) = 1 \lor \raisebox{0.4ex}{\includegraphics[height=1.9ex]{images/number2.png}}~\texttt{shape}(\texttt{v\_2}, \texttt{i}) = 1$
}

\vspace{0.2ex}
{\footnotesize
\raisebox{0.4ex}{\includegraphics[height=1.9ex]{images/number3.png}}~\texttt{else if} $\texttt{ndim}(\texttt{v\_1}) > \texttt{ndim}(\texttt{v\_2})\ \texttt{then}\ \forall\ \texttt{i} \in [0, \texttt{ndim}(\texttt{v\_2}){-}1] :$\\
\hspace*{2.35em}$\texttt{shape}(\texttt{v\_1}, \texttt{ndim}(\texttt{v\_1}){-}\texttt{ndim}(\texttt{v\_2}){+}\texttt{i}) = \texttt{shape}(\texttt{v\_2}, \texttt{i})$\\
\hspace*{2.35em}$\lor\ \texttt{shape}(\texttt{v\_1}, \texttt{ndim}(\texttt{v\_1}){-}\texttt{ndim}(\texttt{v\_2}){+}\texttt{i}) = 1 \lor \texttt{shape}(\texttt{v\_2}, \texttt{i}) = 1$
}

\vspace{0.2ex}
{\footnotesize
\raisebox{0.4ex}{\includegraphics[height=1.9ex]{images/number3.png}}~\texttt{else}$\, \forall\ \texttt{i} \in [0, \texttt{ndim}(\texttt{v\_1}){-}1] :$\\
\hspace*{2.35em}$\texttt{shape}(\texttt{v\_2}, \texttt{ndim}(\texttt{v\_2}){-}\texttt{ndim}(\texttt{v\_1}){+}\texttt{i}) = \texttt{shape}(\texttt{v\_1}, \texttt{i})$\\
\hspace*{2.35em}$\lor\ \texttt{shape}(\texttt{v\_2}, \texttt{ndim}(\texttt{v\_2}){-}\texttt{ndim}(\texttt{v\_1}){+}\texttt{i}) = 1 \lor \texttt{shape}(\texttt{v\_1}, \texttt{i}) = 1$
}

\vspace{0.9ex}
\textbf{Rule 4.} (Input and other tensors should have compatible sizes to avoid broadcasting errors ..)

{\footnotesize
$\{\texttt{v\_1} : \texttt{tensor}, \texttt{v\_2} : \texttt{tensor}\} \models \texttt{if}\ \texttt{ndim}(\texttt{v\_1}) < \texttt{ndim}(\texttt{v\_2})\ \texttt{then}$\\
\hspace*{1em}$\forall\ \texttt{i} \in [0, \texttt{ndim}(\texttt{v\_1}){-}1] :$\\ 
\hspace*{2.35em}$\big(\texttt{shape}(\texttt{v\_1}, \texttt{i}) = \texttt{shape}(\texttt{v\_2}, \texttt{i} + (\texttt{ndim}(\texttt{v\_2}) - \texttt{ndim}(\texttt{v\_1})))$\\
\hspace*{2.35em}$\lor\ \texttt{shape}(\texttt{v\_1}, \texttt{i}) = 1 \lor \texttt{shape}(\texttt{v\_2}, \texttt{i} + (\texttt{ndim}(\texttt{v\_2}) - \texttt{ndim}(\texttt{v\_1}))) = 1 \big)$ ..
}

\vspace{1pt}
\end{minipage}}
\endgroup
}

%% file: figures/grammar_subfig.tex
{\let\texttt\oldtexttt
\small
\noindent
\begingroup
\setlength{\fboxsep}{0pt}
\setlength{\fboxrule}{0.5pt}
\fcolorbox{black!50}{gray!0}{%
\begin{minipage}{\dimexpr\linewidth-2\fboxsep-2\fboxrule\relax}
\vspace{1pt}
  \hspace{1mm}
  \begin{tabular}{@{}l}
    $\mathcal{R} ::= \{\overline{v : \tau}\} \models e
          \quad \tau ::= \texttt{int} \mid \texttt{float} \mid \texttt{bool} \mid \texttt{dtype} \mid \texttt{str}$ \\[0.2ex]
    \hspace*{12em}$\mid \texttt{tensor} \mid \texttt{list}(\tau) \mid \texttt{tuple}(\tau) \mid \tau \uplus \tau$     \\[1.2ex]
    $e$ ::= $l \mid v_\texttt{prim} \mid f(v_\texttt{tensor}[,e]) \mid v_\texttt{tuple}[\texttt{[}e\texttt{]} \mid \texttt{.len}] \mid e_\texttt{b} \mid (e) \mid e \mathbin{\circleddash} e$ \\[0.4ex]
    \hspace*{2.6em}$\mid \texttt{if}\ e\ \texttt{then}\ e\ [\texttt{else}\ e]$ \\[1.2ex]
    $e_\texttt{b}$ ::= $\texttt{true} \mid \texttt{false} \mid e \mathbin{\circledcirc} e \mid e \wedge e \mid e \vee e \mid \forall v \in \texttt{[}e, e\texttt{]} : e(v)$ \\[0.4ex]
    \hspace*{4.7em}$\mid \exists v \in \texttt{[}e, e\texttt{]} : e(v)$ \\[1.4ex]
    \multicolumn{1}{l}{\hspace{2.5em}$f \in \{\texttt{ndim}, \texttt{shape}, \texttt{dtype\_}, \texttt{min}, \texttt{max}\},$} \\[0.6ex]
    \multicolumn{1}{l}{\hspace{2.5em}$\mathbin{\circleddash} \in \{+, -, \times, /\}, \hspace{1em} \mathbin{\circledcirc} \in \{=, \neq, >, <, \geq, \leq\},$} \\[0.6ex]
    \multicolumn{1}{l}{\hspace{2.5em}$l \in \mathbb{R} \cup \texttt{Bool} \cup \texttt{String}, \hspace{1em} v \in \text{Var} = \{\texttt{i}, \texttt{v}_1, \texttt{v}_2, \ldots\}$} \\[1.2ex]
    \multicolumn{1}{l}{\hspace{.5em}($v_\texttt{prim}:$ primitive, \, $v_\texttt{tuple}$: tuple/list, \, $v_\texttt{tensor}$: tensor variables)}
  \end{tabular}
\vspace{1pt}
\end{minipage}}
\endgroup
}

%% file: figures/typing_rules.tex
{\let\texttt\oldtexttt
{\fontsize{7pt}{7pt}\selectfont
\noindent
\begingroup
\setlength{\fboxsep}{0pt}
\setlength{\fboxrule}{0.5pt}
\fcolorbox{black!50}{gray!0}{%
\begin{minipage}{\dimexpr\linewidth-2\fboxsep-2\fboxrule\relax}
\vspace{1pt}
\begin{flushleft}

\[
\frac{
  \Gamma(v) \in \{\texttt{int}, \texttt{float}, \texttt{bool}, \texttt{str}, \texttt{dtype}\}
}{
  \Gamma \vdash v : \Gamma(v)
}
{\quad \small\text{[T-PrimAccess]}}
\]
\vspace{.35ex}
\[
\frac{
  \Gamma(v) = \tau_1 \uplus \tau_2, \quad \tau_1, \tau_2 \in \{\texttt{int}, \texttt{float}, \texttt{bool}, \texttt{str}, \texttt{dtype}\}
}{
  \Gamma \vdash v : \tau_1 \uplus \tau_2
}
{\quad \small\text{[T-UnionAccess]}}
\]
\vspace{.35ex}
\[
\frac{
  \Gamma(v) = \texttt{tuple}(\tau) \text{ or } \texttt{list}(\tau), \quad \Gamma(e) = \texttt{int}
}{
  \Gamma \vdash v[[e]] : \tau
}
{\quad \small\text{[T-TupleAccess]}}
\]
\vspace{.35ex}
\[
\frac{
  \Gamma(v) = \texttt{tuple}(\tau) \text{ or } \texttt{list}(\tau)
}{
  \Gamma \vdash v.\texttt{len} : \texttt{int}
}
{\quad \small\text{[T-TupleLen]}}
\]
\vspace{.35ex}
\[
\frac{
  \Gamma(v) = \texttt{tensor}, \quad f \in \{\texttt{ndim}, \texttt{dtype\_}, \texttt{min}, \texttt{max}\}
}{
  \Gamma \vdash f(v) : \texttt{int}
}
{\quad \small\text{[T-FuncCall-0]}}
\]
\vspace{.35ex}
\[
\frac{
  \Gamma(v) = \texttt{tensor}, \quad \Gamma(e) = \texttt{int}, \quad f = \texttt{shape}
}{
  \Gamma \vdash f(v, e) : \texttt{int}
}
{\quad \small\text{[T-FuncCall-1]}}
\]

\end{flushleft}
\vspace{1pt}
\end{minipage}}
\endgroup
}}

%% file: algorithms/rule_generation.tex
\begin{figure}[t]
  \vspace{-5mm}
  \centering
  \captionsetup{labelformat=empty}
  \begin{algorithm}[H]
    \small    
  \caption{Grammar-based Rule Generation}
  \label{alg:rulegen}
  
  \begin{algorithmic}[1]
    \item[\hspace{1.7em}\textbf{Inputs:}] $a$: API, $\mathcal{G}$: grammar, $\mathcal{R}_\texttt{ex}$: example rules,
    \item[\hspace{4.9em}] $\mathcal{M}$: mutators to augment inputs, $T$: max num. of trials
    \item[\hspace{1.7em}\textbf{Output:}] $\mathcal{R}_a$: ruleset for API $a$
    
    \STATE \textbf{procedure} \textsc{GenerateRules}
    \begin{ALC@g}
      \STATE $\mathcal{R}_a \gets \emptyset$, $c \gets 0$, $F \gets \epsilon$ \COMMENT{ruleset, failure count, feedback}
      \STATE $\mathcal{E}_a \gets \textsc{CollectErrors}(a, \mathcal{M})$
      \WHILE{$c < T$ \textbf{and not} timed out}
          \STATE $r \gets$ \textsc{LLM}(\texttt{PROMPT\_RULE\_GEN}, $\mathcal{G}, F, \texttt{DOC}(a), \mathcal{E}_a, \mathcal{R}_\texttt{ex}$) 
          \STATE $\textit{error} \gets \textsc{Check}(r)$ \label{algo:generaterules:check} 
          \IF{$\textit{error} \neq \emptyset$}
            \STATE $F \gets$ \textit{error}; $c \gets c + 1$ \label{algo:generaterules:failure}
          \ELSE
            \STATE $\mathcal{R}_a \gets \mathcal{R}_a \cup \{r\}$ \label{algo:generaterules:success}
          \ENDIF
      \ENDWHILE
      \STATE \textbf{return} $\mathcal{R}_a$
    \end{ALC@g}
  \end{algorithmic}

  \vspace{0.5em}

  \begin{algorithmic}[1]
    \item[\hspace{1.7em}\textbf{Inputs:}] $a$: API, $\mathcal{M}$: mutators to augment inputs
    \item[\hspace{1.7em}\textbf{Output:}] $\mathcal{E}$: error messages for API $a$
    
    \STATE \textbf{procedure} \textsc{CollectErrors}
    \begin{ALC@g}
      \STATE $\mathcal{E} \gets \emptyset$ 
      \STATE $V_a^{\text{valid}} \gets \textsc{LLM}(\texttt{PROMPT\_INPUT\_GEN}, a, \texttt{DOC}(a))$ \label{algo:collecterrors:valid}
      \STATE $V_a^{\text{mut}} \gets \textsc{Mutate}(V_a^{\text{valid}}, \mathcal{M}, a)$  \label{algo:collecterrors:mutate}
      \STATE $V_a^{\text{rand}} \gets \textsc{Random\_Invalid}(a)$ \label{algo:collecterrors:rand}
      
      \FOR{$v \in V_a^{\text{mut}} \cup V_a^{\text{rand}}$} \label{algo:collecterrors:errorstart}
        \STATE $\textit{outputs}, \textit{error} \gets \textsc{Run}(a,v)$
        \IF{$\textit{error} \neq \emptyset$}
          \STATE $\mathcal{E} \gets \mathcal{E} \cup \{\textit{error}\}$
        \ENDIF
      \ENDFOR \label{algo:collecterrors:errorend}
      \STATE \textbf{return} $\mathcal{E}$
    \end{ALC@g}
  \end{algorithmic}

  \end{algorithm}
  \vspace{-7ex}
\end{figure}

%% file: technique.tex
\section{\tname}



\tname is a neuro-symbolic API-level fuzzer for DL libraries that takes an API
as input and returns a set of potential bug-revealing inputs for that API.
\tname\ uses 1) LLMs for grammar-based constraint learning (the neural part)
and 2) SMT solvers for valid test input generation by solving learned
constraints per API (the symbolic part).

\sloppy
Figure~\ref{fig:workflow} shows the workflow of \tname. 
The technique has three stages: Rule
Generation~(\textsection~\ref{tech:rulegen}), \Invariant
Learning~(\textsection~\ref{tech:refine}), and
Fuzzing~(\textsection~\ref{tech:fuzz}). In the \underline{Rule Generation}
phase, \tname\ uses an LLM to generate a set of candidate rules for a given API
based on a novel grammar representing first-order logic formulae over
API input parameters. In the \underline{\Invariant Learning} phase, \tname\ 
learns likely \invariants by testing the candidate rules on a set of 
valid API inputs. Each rule represents a first-order logic formula, whereas a
\invariant is an instantiation of the formula with (a subset of) API parameters.
In the \underline{Fuzzing} phase, \tname\ executes the API on inputs obtained by
solving the learned \invariants using an SMT solver. \tname\ monitors the
executions for likely bugs (such as crashes or inconsistent outputs) and
reports them to the user.
\subsection{Rule Generation}\label{tech:rulegen}

In the rule generation phase, \tname\ uses the domain knowledge of LLMs 
for valid \dl\ library API usages. More precisely, 
\tname prompts an LLM with a novel grammar to generate candidate
rules for a given API. The grammar represents first-order logic formulae
that capture constraints over API input parameters.
%
Sections~\ref{tech:grammar}
and~\ref{sec:rule-generation-pseudocode} describe, respectively, the
grammar and the rule-generation algorithm.

\subsubsection{Grammar for API Input Constraints}\label{tech:grammar}
\enlargethispage{\baselineskip}

To constrain the responses \added{of the LLM} and extract
information more precisely, \tname provides a grammar \added{of API
  constraints} 
  to the LLM. Figure~\ref{fig:grammar} shows the grammar
that \tname\ uses. 
The grammar can express first-order logic formulae over multiple input
parameters. It is designed to capture relational rules of diverse kinds. A rule
$\mathcal{R}$ consists of a set of variable bindings $\{\overline{v : \tau}\}$
and an expression $e$ over those variables. The variable bindings specify to
which parameter types the rule applies. For example, the following rule
specifies that the second parameter of the API ($v_2$), of integer type, should
be a valid dimension of the first parameter ($v_1$), of tensor type:
$\{v_1\!:\!\texttt{tensor},\, v_2\!:\!\texttt{int}\} \models  
(-1 \times \texttt{ndim}(v_1) \leq v_2) \wedge (v_2 \leq \texttt{ndim}(v_1) - 1)$ Note that \dl\ libraries like PyTorch allows negative 
indexing~\cite{pytorchdocs}. 
%
The grammar supports primitive types, compound types, and union of
types ($\tau\uplus\tau$). The union type is necessary to specify 
constraints for API inputs that admit multiple data types, e.g., 
\texttt{torch.clamp} allows both ``number or tensor'' types for its
parameters~\cite{pytorch_clamp}.
%
The grammar can express properties for tuple and list types, including
indexed access and length. For simplicity, $v_\texttt{tuple}$ denotes
variables of such types. To support tensor-related properties, the
grammar uses five functions: \texttt{ndim} denoting number of
dimensions in tensor, \texttt{shape} denoting the shape of the tensor
(\eg{}, \CodeIn{(3,1,1)}), \texttt{dtype\_} denoting the
library-specific data type (\eg{}, \texttt{float32} in \torch),
\texttt{min}, and \texttt{max}. Finally, the grammar supports 
first-order logical operators \added{(e.g., quantifiers), 
arithmetic operators, and conditionals.}

\subsubsection{Algorithm}
\label{sec:rule-generation-pseudocode}


Algorithm~\ref{alg:rulegen} shows how \tname\ prompts an LLM to generate
candidate rules for a given API. The algorithm takes as input an API $a$, a
grammar $\mathcal{G}$ (\textsection~\ref{tech:grammar}), example rules in the
grammar $\mathcal{R}_\text{ex}$, a set of mutators $\mathcal{M}$, and the number
of trials $T$ for rule generation. It returns candidate rules for the given API as
output.

The utility function \textsc{CollectErrors} finds a set of \emph{runtime}
errors that can be observed by running the API on invalid inputs.
The invalid inputs are generated by 1) querying LLM for valid inputs
($V_a^{\text{valid}}$) (Line~\ref{algo:collecterrors:valid})
and then applying nine kinds of mutations ($V_a^{\text{mut}}$).
For example, we convert to empty tensors or some of the values to
zero or other int/float values (Line~\ref{algo:collecterrors:mutate}),
and 2) generating random values ($V_a^{\text{rand}}$)
(Line~\ref{algo:collecterrors:rand}). The random generation was conducted
with 30 seconds per API. In total, we curate a database of 2,642
and 2,822 error messages for 759 PyTorch APIs and 718 TensorFlow APIs,
respectively. 
In addition, we use few-shot learning~\cite{brown2020language} to improve
performance of the LLM~\cite{Ma2025SpecGen_ICSE}; we manually
define example rules that span every parameter type and property
at least once, and provide to the LLM as few-shot examples.
\input{algorithms/invariant_inference.tex}

The function \textsc{GenerateRules} is responsible for generating rules for the
input API $a$. It does so by iteratively querying an LLM with adjusted prompts
until it reaches the maximum number of failures or it times out. We use 100 as
the failure bound and a one-minute timeout in our evaluation.
\tname provides the grammar, API
documentation, error messages, and example rules to the LLM. \tname includes all
available error messages of the API in a database, which contains an average of
2.54 error messages per API. For each iteration, \tname adjusts the LLM prompt
with one of five types of feedback
(Line~\ref{algo:generaterules:check},\ref{algo:generaterules:failure}):~1)~format
error (e.g., no rule description),~2)~redundant variable bindings,~3)~duplicated
rule,~4)~parsing error, or~5)~success. If successful, the rule is added to the
set of candidate rules (Line~\ref{algo:generaterules:success}).
When given a one-minute time budget, the average number of iterations is 18.4.
Notably, \tname improves efficiency by asking the LLM to return
multiple rules in a single iteration. This enables the generation of 93.37 rules 
on average (max: 213) per API within a minute.

\mypara{Syntactic Validation}~ Our grammar uses a parser implemented
in Lark~\cite{LarkParser}. So, \tname precisely validates the
\emph{syntactic} correctness of LLM-generated rules. Also, \tname
checks for type errors where operators and expected variable types do
not match.
For example, a rule can define tensor-specific operation on primitive type
variables. Figure~\ref{fig:typing_rules} shows the typing rules that \tname\
uses to discard the rules with type errors. We implement the type checking as an
extension of the Lark Transformer~\cite{LarkParser}. 

\subsection{\Invariant Learning}\label{tech:refine} Even if a rule is
grammatically correct with no type error, it should be checked 1) if the rule
is an actual constraint; LLM can generate \emph{semantically}-incorrect rules,
and 2) which API parameters the rule applies to. We call a rule applied to
specific API parameter(s), a \emph{\invariant}. Algorithm~\ref{alg:invlearn}
shows how \tname learns \invariants $\mathcal{I}_a$ for the input API $a$
(\textsc{Learn\Invariants{}}). If all valid values of some parameter(s) 
satisfy a rule, it is likely to be a \invariant~\cite{ErnstPGMPTX2007}.
So, we implement a translator which automatically converts the
grammar-based rule definition to Python code. The code encodes 
constraints as Z3 formulae~\cite{demoura2008z3} and checks 
satisfiability. 
Then, \tname checks the number of parameters ($k$) in the rule
definition. For each ordered subset of $k$ parameters, \tname runs the satisfiability
check by supplying valid values for the parameters. If the rule is satisfied with
all the values, \tname adds it as a \invariant.

\begin{definition}[\Invariant]\label{def:invariant} Let $a$ be an API with
parameter set $\textsc{params}(a)$, and let $r$ be a rule with $k$ variables. We
say that $(r, P)$ is an \emph{\invariant} for $a$, where $P = (p_1, \ldots,
p_k)$ is an ordered list of parameters in $\textsc{params}(a)$, if the rule $r$
holds for all valid inputs of $a$, when the variables in $r$ are assigned to the
corresponding values of $p_1, \ldots, p_k$: $\forall v \in
V_a^{\text{valid}}~.~r(v[p_1], \ldots, v[p_k])$.
\end{definition}



In the set of learned \invariants, there can be redundant ones. For example, a
\invariant can imply another, and LLM can generate semantically-equivalent rules
in different representations.
Redundant \invariants can over-constrain the input space and 
potentially limit the effectiveness of fuzzer. However, exhaustively checking all
redundant rules can be prohibitively expensive (for instance, by checking for
implication among all pairs of \invariants). So, we implement a heuristic to
iteratively eliminate redundant \invariants.

\tname iteratively refines the set $\mathcal{I}_a^{\text{final}}$. \tname first
generates inputs and computes the validity ratio ($v_{\text{orig}}$)
(Line~\ref{algo:invlearn:origstart}--\ref{algo:invlearn:origend}). As we explain
in Sec.~\ref{tech:fuzz}, the translated Python code not only enables
satisfiability checks but also the generation of satisfiable values. Then,
\tname removes each \invariant, one at a time, while keeping all the others
($\mathcal{I}_a^{\text{test}}$), and checks the validity ratio without the
\invariant ($v_{\text{test}}$) (Line~\ref{algo:invlearn:testconcretize}). If the
ratio decreases, it likely implies that the \invariant is an essential one. So,
\tname keeps it in the final \invariant set (Line~\ref{algo:invlearn:final}),
otherwise, discards it (Line~\ref{algo:invlearn:redundant}). Finally, \tname
returns the refined set of \invariants (Line~\ref{algo:invlearn:return}).
\added{This heuristic-based refinement is fully automated and can be easily
applied to any library and API.}

\input{algorithms/fuzz_phase.tex}

\subsection{Fuzzing}\label{tech:fuzz} 
\added{After \tname learns \invariants, the fuzzing phase begins}. 
For efficiency, instead of concrete inputs directly,
\tname generates \emph{\cmodels} using Z3 and saves them in a corpus. 
In each fuzzing iteration, \tname randomly selects an \cmodel, 
generates a concrete input, and executes the target API with the concrete input.

\subsubsection{\Cmodel{} Generation}\label{tech:model} Intuitively, an 
\emph{\cmodel} defines an input sub-space that satisfies all the \invariants,
from which multiple concrete inputs can be generated. For example, if an API has
a single \invariant $\{ v_1: \texttt{tensor} \} \models \texttt{ndim}(v_1)=1$,
the \cmodel can generate any one-dimensional tensor.
Algorithm~\ref{alg:fuzz} shows how \tname generates a corpus of \cmodels{}:
$\mathcal{C}(a)$ for the input API $a$ (\textsc{GenerateModels}). 
User can set a timeout and sampling ratio $p$. 
For our experiments, we use an 1-hour timeout and sampling ratio of 0.3.

\tname first creates a constraint set ($Z$) and adds initial global constraints,
which are universal across all APIs. Examples include the maximum numbers of
dimensions and their sizes (to suppress unrealistic tensors) and maximum size of
tensors in bytes (to avoid memory issue). Also, we constrain the data types and
string values to be chosen from predefined sets. 
Then, \tname checks each \invariant and adds the corresponding logical 
constraints. This process can be easily done due to how our Python 
code is designed: it reuses the logical constraints used during
constraint learning phase. 
\tname first obtains an \cmodel $\mu$ that satisfies the 
accumulated constraints by SMT solving. Since an \cmodel is an abstraction, 
it has a chance to be invalid when concretized; the concretization
involves randomness (\textsection~\ref{tech:concinp}) or the constraints may be 
incomplete. 
So, \tname generates a concrete input, and if the input is valid, adds $\mu$
to the corpus. This process is repeated until the counter $n$ reaches the target
number or the process reaches timeout. The \cmodels are serialized into JSON
format to be stored in and efficiently loaded from the corpus during fuzzing.
\begin{definition}[\Cmodel{}]
Let \(a\) be an API with \invariant set \(\mathcal{I}_a = \{(r_i,
P_i)\}_{i=1}^n\), where each \(r_i(P_i)\) denotes the Z3 
\added{formula} obtained by applying rule \(r_i\) to an ordered parameter subset
\(P_i \subseteq \textsc{params}(a)\). A model \(\mu\) is a total
assignment to the symbolic variables referenced in the formulas
\(\{r_i(P_i)\}\), such that: $\forall (r_i, P_i) \in \mathcal{I}_a~.~r_i(P_i)(\mu).$
\end{definition}

\vspace{-3ex}
\begin{figure}[h!]
  \centering
  \includegraphics[width=0.8\linewidth,trim={0 0.7cm 0
      0},clip]{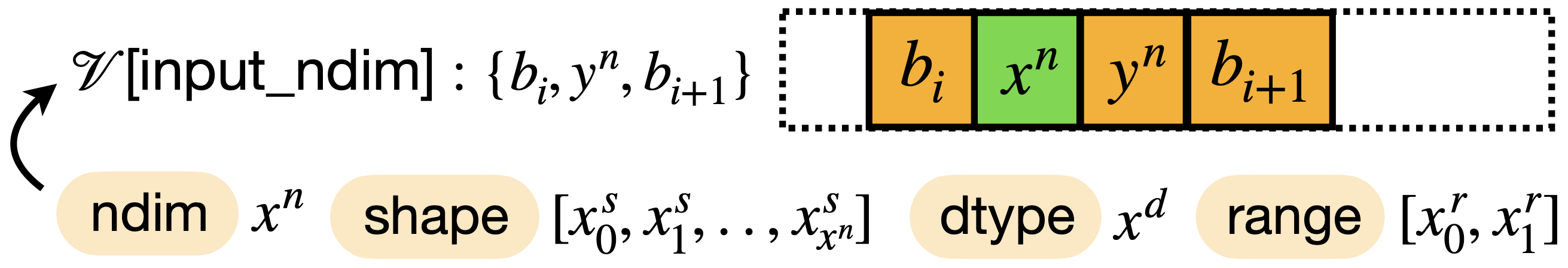}
  \vspace{-1mm}  
  \caption{Example abstract input.}
  \label{fig:example:abstract_input}
  \vspace{-3ex}
\end{figure}

\mypara{Optimizations}
By default, Z3 returns the same solution (i.e., \cmodel for \tname) given the
same \invariants. To improve the input diversity, we develop two tactics.
\textbf{(1)~Blocking:} \tname adds \emph{blocking} constraints to
prevent the same value in the subsequent \cmodels. For every \cmodel generated,
\tname records the value of every Z3 variable ($\mathcal{V}$). It samples a
subset of the values (using a sampling ratio $p$), and adds constraints that
block the corresponding variables from being assigned the same values in future
iterations (Algorithm~\ref{alg:fuzz}, Line~\ref{algo:absinput:blockstart}).
\textbf{(2)~Bucketing:} Even with the blocking mechanism, Z3 often returns
values closer to the limits specified in the constraint, which limits diversity
of inputs (such as tensors). So, to further improve input diversity, we
implement a bucketing technique in \tname. For each data type (integers,
floating point, tensors, and booleans), we partition the input space into
buckets $[b_1, b_2, \ldots, b_k]$, and add additional constraints $\phi: b_i \le
v \le b_{i+1}$ that limit the values of a variable $v$ of that type. This
approach allows \tname to uniformly explore the input space.
Lines~\ref{algo:absinput:bucketstart} and~\ref{algo:absinput:bucketend} in
Algorithm~\ref{alg:fuzz} show how \tname implements bucketing.


Figure~\ref{fig:example:abstract_input} shows an example abstract input for an
\texttt{input} tensor. 
A tensor is abstracted based on dimension, shape, data
type, and range properties. For the dimension property, a bucket of size four is
selected (simplified). As three values are blocked from previous abstract
inputs (colored in orange), $x^n$ is chosen for the property.

\input{tables/num_target_apis.tex}

\subsubsection{Concrete Input Generation}\label{tech:concinp}
Once the corpus of \cmodels are constructed, \tname begins fuzzing. \tname
randomly selects an \cmodel from the corpus and constructs a concrete input using
the values assigned to symbolic variables in the \cmodel. These variables
represent structural and semantic properties of input parameters and constrain
the generation of valid inputs. 
For instance, \tname uses the four tensor properties to construct 
a tensor of the specified dimension, shape, and type, while
uniformly sampling its contents from the value range.

This is a repetitive, and notably, the only online phase of \tname: the rule
generation, \invariant learning, and \cmodel generation are all offline phases
which need to be done only once per API. This brings efficiency benefits
compared to \pathfinder~\cite{Kim2025LightweightConcolic} that computes
and solves the path conditions to generate every input.
Also, the inputs generated by \tname have a validity
ratio of near 100\%. \tname's separation of stages enables a systematic and
balanced exploration of input space: \cmodel generation expands the breadth,
while concrete input generation explores depth. As a result, \tname
achieves similar or higher coverage compared to previous approaches in most cases.

%% file: algorithms/invariant_inference.tex
\begin{figure}[t]
  \vspace{-3mm}
  \centering
  \captionsetup{labelformat=empty}
  \begin{algorithm}[H]
    \small
  \caption{\Invariant Learning of \tname}
  \label{alg:invlearn}
  \begin{algorithmic}[1]

    \item[\hspace{1.7em}\textbf{Inputs:}] $\mathcal{R}_a$: rules for API $a$,\, $V_a^{\text{valid}}$: valid inputs for $a$,
    \item[\hspace{4.9em}] $T$: num. of trials
    \item[\hspace{1.7em}\textbf{Output:}] $\mathcal{I}_a$: \invariants for $a$

    \STATE \textbf{procedure} \textsc{Learn\Invariants{}}
      \begin{ALC@g}
      \STATE $\mathcal{I}_a \gets \emptyset$ 
      \FOR{$r \in \mathcal{R}(a)$}
        \IF{$\textsc{TypeCheck}(r) = \bot$}
          \STATE \textbf{continue}
        \ENDIF
        \STATE $k \gets |\textsc{Vars}(r)|$ 
        \FOR{$P \in \{(p_1,\ldots, p_k) | p_i \in \textsc{Params}(a)\}$}
          \IF{$\textsc{Is\Invariant}((r, P), V_a^{\text{valid}})$}
            \STATE $\mathcal{I}_a \gets \mathcal{I}_a \cup \{(r, P)\}$  \COMMENT{See Def.~\ref{def:invariant}}
          \ENDIF
        \ENDFOR
      \ENDFOR

      \STATE $\mathcal{I}_a^{\text{final}} \gets \emptyset$, $v_{\text{orig}} \gets 0$, $I_a^{r} \gets \emptyset$ \label{algo:invlearn:origstart} 
      
      \FOR{$t = 1$ \textbf{to} $T$}
        \STATE $x \gets $ \textsc{Concretize}(a, \textsc{Z3GetModel}($\mathcal{I}_a$)) \label{algo:invlearn:origconcretize} 
        \IF{\textsc{IsValid}($a, x$) } 
            \STATE $v_{\text{orig}} \gets v_{\text{orig}} + 1$ 
        \ENDIF
      \ENDFOR \label{algo:invlearn:origend}
      
      \FOR{$(r, P) \in \mathcal{I}_a$}
        \STATE $\mathcal{I}_a^{\text{test}} \gets \mathcal{I}_a \setminus \{(r, P)\} \setminus I_a^{r}$ 
        \STATE $v_{\text{test}} \gets 0$ 
        \FOR{$t = 1$ \textbf{to} $T$}
          \STATE $x \gets $ \textsc{Concretize}(a, \textsc{Z3GetModel}($\mathcal{I}_a^{\text{test}}$)) \label{algo:invlearn:testconcretize}
          \IF{\textsc{IsValid}($a, x$)} 
            \STATE $v_{\text{test}} \gets v_{\text{test}} + 1$ 
          \ENDIF
        \ENDFOR
        \IF{$v_{\text{test}} < v_{\text{orig}}$}
          \STATE $\mathcal{I}_a^{\text{final}} \gets \mathcal{I}_a^{\text{final}} \cup \{(r, P)\}$ \label{algo:invlearn:final}
        \ELSE
          \STATE $I_a^{r} \gets I_a^{r} \cup \{(r, P)\}$ \label{algo:invlearn:redundant}
        \ENDIF
      \ENDFOR

      \STATE \textbf{return} $\mathcal{I}_a^{\text{final}}$ \label{algo:invlearn:return}
      \end{ALC@g}

  \end{algorithmic}
  \end{algorithm}
  \vspace{-4ex}
\end{figure}

%% file: algorithms/fuzz_phase.tex
\begin{figure}[t]
  \vspace{-2.3mm}
  \centering
  \begin{algorithm}[H]
    \small
    \caption{\Cmodel{} Generation}
    \label{alg:fuzz}
    \begin{algorithmic}[1] 

      \item[\hspace{1.7em}\textbf{Inputs:}] $\mathcal{I}_a$: \invariants for API $a$,\, $N$: num. of \cmodels{},
      \item[\hspace{4.9em}] $p$: sampling ratio
      \item[\hspace{1.7em}\textbf{Output:}] $\mathcal{C}_a$: corpus of \cmodels{} for API $a$

      \STATE \textbf{procedure} \textsc{Generate\CMODEL{}}:
      \begin{ALC@g}
        \STATE $\mathcal{V} \gets \emptyset$, $\mathcal{C}_a \gets \emptyset$ 
        
        \WHILE{not Timeout}
          \STATE $Z \gets \texttt{GLOBAL\_CONSTRAINTS}$
          \FORALL{$(r, P) \in \mathcal{I}_a$} 
            \STATE add $r(P)$ to $Z$ 
          \ENDFOR
          \STATE $S \gets \textsc{Sample}(\textsc{Variables}(Z), p)$
          
          \FORALL{$v_i:\tau_i \in S$}
            \IF{$\mathcal{V}[v_i] \neq \emptyset$}
              \STATE $val \gets \textsc{Sample}(\mathcal{V}[v_i], 1)$
              \STATE $Z \gets Z \cup \Big\{ \{v_i: \tau_i \} \models v_i \neq val \Big\}$ \label{algo:absinput:blockstart}
            \ENDIF
            \STATE $[l, u] \gets \textsc{Sample}(\textsc{Buckets}(\tau_i), 1)$ \label{algo:absinput:bucketstart}
            \STATE $Z \gets Z \cup \Big\{ \{v_i: \tau_i \} \models l \le v_i \le u \Big\}$ \label{algo:absinput:bucketend}
          \ENDFOR
          
          \IF{\textsc{Z3Solver}($Z$) $=$ \texttt{SAT}}
            \STATE $\mu \gets \textsc{Z3GetModel}(Z)$ 
            \STATE $x \gets \textsc{Concretize}(a, \mu)$
            \IF{\textsc{IsValid}($a$, $x$)}
              \FORALL{$v \in \textsc{Variables}(\mu)$} 
                \STATE $\mathcal{V}[v] \gets \mathcal{V}[v] \cup \{\mu[v]\}$ 
              \ENDFOR
              \STATE $\mathcal{C}(a) \gets \mathcal{C}(a) \cup \{\mu\}$
            \ENDIF
          \ENDIF
        \ENDWHILE
        \RETURN $\mathcal{C}(a)$
      \end{ALC@g}
    \end{algorithmic}
  \end{algorithm}
  \vspace{-6ex}
\end{figure}

%% file: tables/num_target_apis.tex
\begin{table}[t!]\centering
\vspace{1ex}
\caption{Number of APIs \tname supports and number of APIs in
  common between \tname\ and each baseline.}\label{tab:num_target_apis}
\vspace{-2ex}  
\resizebox{\linewidth}{!}{%
  \begin{tabular}{lrrrr}
  \toprule
  & \tname & $\cap$ \acetest & $\cap$ \pathfinder & $\cap$ \titanfuzz  \\
  \midrule
  \torch & \nAPIsSL & \nAPIsAC  & \nAPIsPF & \nAPIsTF \\
  \tf  & \nAPIsSLtf & \nAPIsACtf  & \nAPIsPFtf & \nAPIsTFtf \\
  \bottomrule
  \end{tabular}
}
\vspace{-4ex}  
\end{table}

%% file: exp-setup.tex
\input{tables/rq1_manual_analysis.tex}
\input{figures/rq1_validity_bars_table.tex}
\section{Experimental Setup}

\mypara{Implementation Details}
Our tool \tname{} is implemented in Python 3.12. We use the Z3 SMT Solver
\cite{demoura2008z3}, particularly the Python package (z3-solver version 4.14.1)
that allows us to construct the \invariants and to generate inputs by solving
the constraints. We use LLMs to generate the candidate rules, a list of valid
inputs for each API to use for validating the \invariants and to generate the
signatures of the APIs. We use Google Gemini 2.0 Flash \cite{team2023gemini} via
the Google Cloud Platform (GCP) as the LLM in our experiments. For syntactic validation,
of the generated rules, we use Lark version 1.2.2. We also use GPT-5, 
Claude Sonnet 4.5, Gemma 3 27B, and Qwen3-Coder-30B for 
further analysis~(\textsection~\ref{discussion}). 

\mypara{Baselines}
For evaluation, we compare with two state-of-the-art constraint-based
approaches: \acetest{} \cite{Shi_ETAL_ISSTA23} and \pathfinder{}
\cite{pathfinder2025}.
They both use constraints, but unlike \tname which is
based on API parameters, their constraints are path 
conditions for a specific input.
We also include a technique that is not constraint-based, but uses LLMs to
generate inputs: \titanfuzz{} \cite{Deng_ETAL_ISSTA23}. We compare with these
tools based on standard metrics such as coverage and validity ratio.
Table~\ref{tab:num_target_apis} presents the number of APIs we can support along
with the number of APIs common with each \sota{}.

\looseness=-1
\mypara{Experimental Setup}
Our experiments target \torch{} and \tf{}, the most popular DL libraries. We
evaluated \tname{} against three state-of-the-art baselines: \acetest{}
\cite{Shi_ETAL_ISSTA23}, \titanfuzz{} \cite{Deng_ETAL_ISSTA23}, and
\pathfinder{} \cite{Kim2025LightweightConcolic}. While \tname{} is compatible with the
latest library versions (\torchVersion{} and \tfVersion{}), all coverage
comparisons were performed on instrumented versions of \torch{} 2.2.0 and \tf{}
2.16.1 to ensure a fair comparison, as porting the \pathfinder{} artifact is
non-trivial. Table~\ref{tab:num_target_apis} reports the number of APIs we
support and the overlap with each baseline. To ensure a fair and uniform
evaluation, we use llvm-cov~\cite{llvm-cov} to measure branch coverage for all
tools.  For \acetest{} and \titanfuzz{}, we used monkey
patching~\cite{hunt2019monkey} to isolate the coverage of only the target APIs.
For validity ratio computation, we only consider an input valid if it does not
raise an exception within the framework, determined using
scripts from each tool's artifact. 

For bug detection, we use oracles that are standard across the baselines
\cite{Wei2022,Deng_ETAL_ISSTA23,Deng_ETAL_FSE22}. These are 1) crash oracle:
detecting when an input raises a signal (e.g. segmentation fault, aborted)
and 2) differential
oracle: detecting inconsistencies between outputs by running on different
devices (CPU and GPU). \tname logs any oracle violation, and after manually reviewing the
logs, we submit bug reports that are determined by the authors to be true
positives. \added{During the review process, we further classify the
inconsistencies into three categories:~(i) \emph{NaN} when only one platform
generates a ``Not a Number'' value,~(ii) \emph{overflow} when only one platform
encounters an arithmetic overflow, and~(iii) \emph{inconsistent} when there is
numerical inconsistency in the outputs indicating potential bugs in
implementation.} 

All experiments were conducted on an Ubuntu 22.04 server with dual AMD EPYC
9684X CPUs and 768GB of RAM. To ensure consistency, we run techniques in Docker
containers. We allocate a 180-second time budget per API per tool. 
We use 30 and 0.3 for the number of trials and sampling ratio in 
Algorithms~\ref{alg:invlearn}--\ref{alg:fuzz}, respectively. 

\Space{we standardized our methodology. Branch
coverage for all tools was measured using llvm-cov~\cite{llvm-cov} on libraries
instrumented with LLVM's coverage flags. For \acetest{} and \titanfuzz{}, we
used monkey patching~\cite{hunt2019monkey} to isolate the coverage of only the
target APIs. For \pathfinder{}, we modified its artifact to use llvm-cov instead
of its default LCOV~\cite{lcov} tool. Finally, we used a consistent definition
of input validity across all tools: an input is considered valid if its
execution does not raise an exception within the framework, determined using
scripts from each tool's artifact.}

%% file: tables/rq1_manual_analysis.tex
\begin{table*}[t]
\centering
\footnotesize
\caption{Comparison of constraints against ground truth (\#E: num. of error messages, $V^R$: validity ratio of random generation, $I^{G}$ and $I^{S}$: \invariants of ground truth and \tname, $M^{G}$ and $M^{S}$: avg. model generation time of ground truth and \tname).}
\label{tab:rq1-inv}
\setlength{\tabcolsep}{2.1pt}
\renewcommand{\arraystretch}{0.8} 
\vspace{-3ex}
\begin{tabular}{@{}lccccccclccccccc@{}}
\multicolumn{8}{c}{\textbf{PyTorch}} & \multicolumn{8}{c}{\textbf{TensorFlow}} \\[.3ex]
\toprule
\multicolumn{1}{c}{\textbf{API}} & \multicolumn{1}{c}{\textbf{\#E}} & \multicolumn{1}{c}{\textbf{$\boldsymbol{V^R}$}} & \multicolumn{1}{c}{\textbf{$\boldsymbol{I^{G}}$ (Recall \%)}} & \multicolumn{1}{c}{\textbf{$\boldsymbol{M^{G}}$(s)}} & \multicolumn{1}{c}{\textbf{$\boldsymbol{I^{S}}$ (Precision \%)}} & \multicolumn{1}{c}{\textbf{$\boldsymbol{M^{S}}$(s)}} & \multicolumn{1}{c}{\textbf{}} & \multicolumn{1}{c}{\textbf{API}} & \multicolumn{1}{c}{\textbf{\#E}} & \multicolumn{1}{c}{\textbf{$\boldsymbol{V^R}$}} & \multicolumn{1}{c}{\textbf{$\boldsymbol{I^{G}}$ (Recall \%)}} & \multicolumn{1}{c}{\textbf{$\boldsymbol{M^{G}}$(s)}} & \multicolumn{1}{c}{\textbf{$\boldsymbol{I^{S}}$ (Precision \%)}} & \multicolumn{1}{c}{\textbf{$\boldsymbol{M^{S}}$(s)}} & \multicolumn{1}{c}{\textbf{}} \\
\midrule
\multicolumn{1}{c}{\textit{abs}} & \multicolumn{1}{c}{\torchabsE} & \multicolumn{1}{c}{\torchabsVR} & \multicolumn{1}{c}{2/2 (100.0\%)$^{\scriptscriptstyle\dagger}$} & \multicolumn{1}{c}{\torchabsMG} & \multicolumn{1}{c}{3/3 (100.0\%)$^{\scriptscriptstyle\dagger}$} & \multicolumn{1}{c}{\torchabsMS} & \multicolumn{1}{c}{} & \multicolumn{1}{c}{\textit{experimental.numpy.-}} & \multicolumn{1}{c}{\tensorflowexperimentalnumpyanyE} & \multicolumn{1}{c}{\tensorflowexperimentalnumpyanyVR} & \multicolumn{1}{c}{1/1 (100.0\%)$^{\scriptscriptstyle\dagger}$} & \multicolumn{1}{c}{\tensorflowexperimentalnumpyanyMG} & \multicolumn{1}{c}{4/5 (80.0\%)} & \multicolumn{1}{c}{\tensorflowexperimentalnumpyanyMS} & \multicolumn{1}{c}{} \\
\multicolumn{1}{c}{\textit{argmax}} & \multicolumn{1}{c}{\torchargmaxE} & \multicolumn{1}{c}{\torchargmaxVR} & \multicolumn{1}{c}{3/3 (100.0\%)$^{\scriptscriptstyle\dagger}$} & \multicolumn{1}{c}{\torchargmaxMG} & \multicolumn{1}{c}{6/6 (100.0\%)$^{\scriptscriptstyle\dagger}$} & \multicolumn{1}{c}{\torchargmaxMS} & \multicolumn{1}{c}{} & \multicolumn{1}{c}{\textit{eye}} & \multicolumn{1}{c}{\tensorfloweyeE} & \multicolumn{1}{c}{\tensorfloweyeVR} & \multicolumn{1}{c}{3/3 (100.0\%)$^{\scriptscriptstyle\dagger}$} & \multicolumn{1}{c}{\tensorfloweyeMG} & \multicolumn{1}{c}{8/8 (100.0\%)$^{\scriptscriptstyle\dagger}$} & \multicolumn{1}{c}{\tensorfloweyeMS} & \multicolumn{1}{c}{} \\
\multicolumn{1}{c}{\textit{channel\_shuffle}} & \multicolumn{1}{c}{\torchchannelshuffleE} & \multicolumn{1}{c}{\torchchannelshuffleVR} & \multicolumn{1}{c}{3/3 (100.0\%)$^{\scriptscriptstyle\dagger}$} & \multicolumn{1}{c}{\torchchannelshuffleMG} & \multicolumn{1}{c}{1/1 (100.0\%)$^{\scriptscriptstyle\dagger}$} & \multicolumn{1}{c}{\torchchannelshuffleMS} & \multicolumn{1}{c}{} & \multicolumn{1}{c}{\textit{image.ssim}} & \multicolumn{1}{c}{\tensorflowimagessimE} & \multicolumn{1}{c}{\tensorflowimagessimVR} & \multicolumn{1}{c}{6/6 (100.0\%)$^{\scriptscriptstyle\dagger}$} & \multicolumn{1}{c}{\tensorflowimagessimMG} & \multicolumn{1}{c}{9/10 (90.0\%)} & \multicolumn{1}{c}{\tensorflowimagessimMS} & \multicolumn{1}{c}{} \\
\multicolumn{1}{c}{\textit{floor}} & \multicolumn{1}{c}{\torchfloorE} & \multicolumn{1}{c}{\torchfloorVR} & \multicolumn{1}{c}{1/3 (33.3\%)} & \multicolumn{1}{c}{\torchfloorMG} & \multicolumn{1}{c}{1/1 (100.0\%)$^{\scriptscriptstyle\dagger}$} & \multicolumn{1}{c}{\torchfloorMS} & \multicolumn{1}{c}{} & \multicolumn{1}{c}{\textit{image.transpose}} & \multicolumn{1}{c}{\tensorflowimagetransposeE} & \multicolumn{1}{c}{\tensorflowimagetransposeVR} & \multicolumn{1}{c}{2/2 (100.0\%)$^{\scriptscriptstyle\dagger}$} & \multicolumn{1}{c}{\tensorflowimagetransposeMG} & \multicolumn{1}{c}{4/4 (100.0\%)$^{\scriptscriptstyle\dagger}$} & \multicolumn{1}{c}{\tensorflowimagetransposeMS} & \multicolumn{1}{c}{} \\
\multicolumn{1}{c}{\textit{lcm}} & \multicolumn{1}{c}{\torchlcmE} & \multicolumn{1}{c}{\torchlcmVR} & \multicolumn{1}{c}{2/2 (100.0\%)$^{\scriptscriptstyle\dagger}$} & \multicolumn{1}{c}{\torchlcmMG} & \multicolumn{1}{c}{9/9 (100.0\%)$^{\scriptscriptstyle\dagger}$} & \multicolumn{1}{c}{\torchlcmMS} & \multicolumn{1}{c}{} & \multicolumn{1}{c}{\textit{linalg.inv}} & \multicolumn{1}{c}{\tensorflowlinalginvE} & \multicolumn{1}{c}{\tensorflowlinalginvVR} & \multicolumn{1}{c}{3/3 (100.0\%)$^{\scriptscriptstyle\dagger}$} & \multicolumn{1}{c}{\tensorflowlinalginvMG} & \multicolumn{1}{c}{13/13 (100.0\%)$^{\scriptscriptstyle\dagger}$} & \multicolumn{1}{c}{\tensorflowlinalginvMS} & \multicolumn{1}{c}{} \\
\multicolumn{1}{c}{\textit{linalg.solve\_ex}} & \multicolumn{1}{c}{\torchlinalgsolveexE} & \multicolumn{1}{c}{\torchlinalgsolveexVR} & \multicolumn{1}{c}{5/6 (83.3\%)} & \multicolumn{1}{c}{\torchlinalgsolveexMG} & \multicolumn{1}{c}{9/9 (100.0\%)$^{\scriptscriptstyle\dagger}$} & \multicolumn{1}{c}{\torchlinalgsolveexMS} & \multicolumn{1}{c}{} & \multicolumn{1}{c}{\textit{math.invert\_permut-}} & \multicolumn{1}{c}{\tensorflowmathinvertpermutationE} & \multicolumn{1}{c}{\tensorflowmathinvertpermutationVR} & \multicolumn{1}{c}{3/3 (100.0\%)$^{\scriptscriptstyle\dagger}$} & \multicolumn{1}{c}{\tensorflowmathinvertpermutationMG} & \multicolumn{1}{c}{6/6 (100.0\%)$^{\scriptscriptstyle\dagger}$} & \multicolumn{1}{c}{\tensorflowmathinvertpermutationMS} & \multicolumn{1}{c}{} \\
\multicolumn{1}{c}{\textit{narrow}} & \multicolumn{1}{c}{\torchnarrowE} & \multicolumn{1}{c}{\torchnarrowVR} & \multicolumn{1}{c}{3/3 (100.0\%)$^{\scriptscriptstyle\dagger}$} & \multicolumn{1}{c}{\torchnarrowMG} & \multicolumn{1}{c}{11/11 (100.0\%)$^{\scriptscriptstyle\dagger}$} & \multicolumn{1}{c}{\torchnarrowMS} & \multicolumn{1}{c}{} & \multicolumn{1}{c}{\textit{math.sigmoid}} & \multicolumn{1}{c}{\tensorflowmathsigmoidE} & \multicolumn{1}{c}{\tensorflowmathsigmoidVR} & \multicolumn{1}{c}{1/1 (100.0\%)$^{\scriptscriptstyle\dagger}$} & \multicolumn{1}{c}{\tensorflowmathsigmoidMG} & \multicolumn{1}{c}{4/4 (100.0\%)$^{\scriptscriptstyle\dagger}$} & \multicolumn{1}{c}{\tensorflowmathsigmoidMS} & \multicolumn{1}{c}{} \\
\multicolumn{1}{c}{\textit{nn.ReLU}} & \multicolumn{1}{c}{\torchnnReLUE} & \multicolumn{1}{c}{\torchnnReLUVR} & \multicolumn{1}{c}{1/1 (100.0\%)$^{\scriptscriptstyle\dagger}$} & \multicolumn{1}{c}{\torchnnReLUMG} & \multicolumn{1}{c}{1/1 (100.0\%)$^{\scriptscriptstyle\dagger}$} & \multicolumn{1}{c}{\torchnnReLUMS} & \multicolumn{1}{c}{} & \multicolumn{1}{c}{\textit{nn.crelu}} & \multicolumn{1}{c}{\tensorflownncreluE} & \multicolumn{1}{c}{\tensorflownncreluVR} & \multicolumn{1}{c}{3/3 (100.0\%)$^{\scriptscriptstyle\dagger}$} & \multicolumn{1}{c}{\tensorflownncreluMG} & \multicolumn{1}{c}{8/10 (80.0\%)} & \multicolumn{1}{c}{\tensorflownncreluMS} & \multicolumn{1}{c}{} \\
\multicolumn{1}{c}{\textit{nn.Softplus}} & \multicolumn{1}{c}{\torchnnSoftplusE} & \multicolumn{1}{c}{\torchnnSoftplusVR} & \multicolumn{1}{c}{1/1 (100.0\%)$^{\scriptscriptstyle\dagger}$} & \multicolumn{1}{c}{\torchnnSoftplusMG} & \multicolumn{1}{c}{4/4 (100.0\%)$^{\scriptscriptstyle\dagger}$} & \multicolumn{1}{c}{\torchnnSoftplusMS} & \multicolumn{1}{c}{} & \multicolumn{1}{c}{\textit{nn.isotonic\_regres-}} & \multicolumn{1}{c}{\tensorflownnisotonicregressionE} & \multicolumn{1}{c}{\tensorflownnisotonicregressionVR} & \multicolumn{1}{c}{1/1 (100.0\%)$^{\scriptscriptstyle\dagger}$} & \multicolumn{1}{c}{\tensorflownnisotonicregressionMG} & \multicolumn{1}{c}{3/3 (100.0\%)$^{\scriptscriptstyle\dagger}$} & \multicolumn{1}{c}{\tensorflownnisotonicregressionMS} & \multicolumn{1}{c}{} \\
\multicolumn{1}{c}{\textit{nn.functional.alpha-}} & \multicolumn{1}{c}{\torchnnfunctionalalphadropoutE} & \multicolumn{1}{c}{\torchnnfunctionalalphadropoutVR} & \multicolumn{1}{c}{2/2 (100.0\%)$^{\scriptscriptstyle\dagger}$} & \multicolumn{1}{c}{\torchnnfunctionalalphadropoutMG} & \multicolumn{1}{c}{7/7 (100.0\%)$^{\scriptscriptstyle\dagger}$} & \multicolumn{1}{c}{\torchnnfunctionalalphadropoutMS} & \multicolumn{1}{c}{} & \multicolumn{1}{c}{\textit{signal.fft}} & \multicolumn{1}{c}{\tensorflowsignalfftE} & \multicolumn{1}{c}{\tensorflowsignalfftVR} & \multicolumn{1}{c}{2/2 (100.0\%)$^{\scriptscriptstyle\dagger}$} & \multicolumn{1}{c}{\tensorflowsignalfftMG} & \multicolumn{1}{c}{2/3 (66.7\%)} & \multicolumn{1}{c}{\tensorflowsignalfftMS} & \multicolumn{1}{c}{} \\
\multicolumn{1}{c}{\added{\textit{nn.init.sparse\_}}} & \multicolumn{1}{c}{\added{\torchnninitsparseE}} & \multicolumn{1}{c}{\added{\torchnninitsparseVR}} & \multicolumn{1}{c}{\added{2/2 (100.0\%)$^{\scriptscriptstyle\dagger}$}} & \multicolumn{1}{c}{\added{\torchnninitsparseMG}} & \multicolumn{1}{c}{\added{3/4 (75.0\%)}} & \multicolumn{1}{c}{\added{\torchnninitsparseMS}} & \multicolumn{1}{c}{} & \multicolumn{1}{c}{\added{\textit{linalg.matrix\_rank}}} & \multicolumn{1}{c}{\added{\tensorflowlinalgmatrixrankE}} & \multicolumn{1}{c}{\added{\tensorflowlinalgmatrixrankVR}} & \multicolumn{1}{c}{\added{1/1 (100.0\%)$^{\scriptscriptstyle\dagger}$}} & \multicolumn{1}{c}{\added{\tensorflowlinalgmatrixrankMG}} & \multicolumn{1}{c}{\added{1/1 (100.0\%)$^{\scriptscriptstyle\dagger}$}} & \multicolumn{1}{c}{\added{\tensorflowlinalgmatrixrankMS}} & \multicolumn{1}{c}{} \\
\multicolumn{1}{c}{\added{\textit{linalg.householder\-}}} & \multicolumn{1}{c}{\added{\torchlinalghouseholderproductE}} & \multicolumn{1}{c}{\added{\torchlinalghouseholderproductVR}} & \multicolumn{1}{c}{\added{3/4 (75.0\%)}} & \multicolumn{1}{c}{\added{\torchlinalghouseholderproductMG}} & \multicolumn{1}{c}{\added{12/12 (100.0\%)$^{\scriptscriptstyle\dagger}$}} & \multicolumn{1}{c}{\added{\torchlinalghouseholderproductMS}} & \multicolumn{1}{c}{} & \multicolumn{1}{c}{\added{\textit{sparse.eye}}} & \multicolumn{1}{c}{\added{\tensorflowsparseeyeE}} & \multicolumn{1}{c}{\added{\tensorflowsparseeyeVR}} & \multicolumn{1}{c}{\added{2/2 (100.0\%)$^{\scriptscriptstyle\dagger}$}} & \multicolumn{1}{c}{\added{\tensorflowsparseeyeMG}} & \multicolumn{1}{c}{\added{4/4 (100.0\%)$^{\scriptscriptstyle\dagger}$}} & \multicolumn{1}{c}{\added{\tensorflowsparseeyeMS}} & \multicolumn{1}{c}{} \\
\multicolumn{1}{c}{\added{\textit{nn.init.eye\_}}} & \multicolumn{1}{c}{\added{\torchnniniteyeE}} & \multicolumn{1}{c}{\added{\torchnniniteyeVR}} & \multicolumn{1}{c}{\added{1/1 (100.0\%)$^{\scriptscriptstyle\dagger}$}} & \multicolumn{1}{c}{\added{\torchnniniteyeMG}} & \multicolumn{1}{c}{\added{4/4 (100.0\%)$^{\scriptscriptstyle\dagger}$}} & \multicolumn{1}{c}{\added{\torchnniniteyeMS}} & \multicolumn{1}{c}{} & \multicolumn{1}{c}{\added{\textit{make\_tensor\_proto}}} & \multicolumn{1}{c}{\added{\tensorflowmaketensorprotoE}} & \multicolumn{1}{c}{\added{\tensorflowmaketensorprotoVR}} & \multicolumn{1}{c}{\added{2/2 (100.0\%)$^{\scriptscriptstyle\dagger}$}} & \multicolumn{1}{c}{\added{\tensorflowmaketensorprotoMG}} & \multicolumn{1}{c}{\added{5/6 (83.3\%)}} & \multicolumn{1}{c}{\added{\tensorflowmaketensorprotoMS}} & \multicolumn{1}{c}{} \\
\multicolumn{1}{c}{\added{\textit{view\_as\_real}}} & \multicolumn{1}{c}{\added{\torchviewasrealE}} & \multicolumn{1}{c}{\added{\torchviewasrealVR}} & \multicolumn{1}{c}{\added{1/1 (100.0\%)$^{\scriptscriptstyle\dagger}$}} & \multicolumn{1}{c}{\added{\torchviewasrealMG}} & \multicolumn{1}{c}{\added{7/7 (100.0\%)$^{\scriptscriptstyle\dagger}$}} & \multicolumn{1}{c}{\added{\torchviewasrealMS}} & \multicolumn{1}{c}{} & \multicolumn{1}{c}{\added{\textit{unique}}} & \multicolumn{1}{c}{\added{\tensorflowuniqueE}} & \multicolumn{1}{c}{\added{\tensorflowuniqueVR}} & \multicolumn{1}{c}{\added{1/1 (100.0\%)$^{\scriptscriptstyle\dagger}$}} & \multicolumn{1}{c}{\added{\tensorflowuniqueMG}} & \multicolumn{1}{c}{\added{5/6 (83.3\%)}} & \multicolumn{1}{c}{\added{\tensorflowuniqueMS}} & \multicolumn{1}{c}{} \\
\multicolumn{1}{c}{\added{\textit{special.expit}}} & \multicolumn{1}{c}{\added{\torchspecialexpitE}} & \multicolumn{1}{c}{\added{\torchspecialexpitVR}} & \multicolumn{1}{c}{\added{1/1 (100.0\%)$^{\scriptscriptstyle\dagger}$}} & \multicolumn{1}{c}{\added{\torchspecialexpitMG}} & \multicolumn{1}{c}{\added{1/1 (100.0\%)$^{\scriptscriptstyle\dagger}$}} & \multicolumn{1}{c}{\added{\torchspecialexpitMS}} & \multicolumn{1}{c}{} & \multicolumn{1}{c}{\added{\textit{unique\_with\_counts}}} & \multicolumn{1}{c}{\added{\tensorflowuniquewithcountsE}} & \multicolumn{1}{c}{\added{\tensorflowuniquewithcountsVR}} & \multicolumn{1}{c}{\added{1/1 (100.0\%)$^{\scriptscriptstyle\dagger}$}} & \multicolumn{1}{c}{\added{\tensorflowuniquewithcountsMG}} & \multicolumn{1}{c}{\added{3/5 (60.0\%)}} & \multicolumn{1}{c}{\added{\tensorflowuniquewithcountsMS}} & \multicolumn{1}{c}{} \\
\midrule
\removeRow{\multicolumn{1}{c}{\textbf{\deleted{Total}}} & \multicolumn{1}{c}{\deleted{39}} & \multicolumn{1}{c}{\deleted{-}} & \multicolumn{1}{c}{\deleted{\textbf{21/24 (87.5\%)}}} & \multicolumn{1}{c}{\deleted{2.10}} & \multicolumn{1}{c}{\deleted{\textbf{52/53 (98.1\%)}}} & \multicolumn{1}{c}{\deleted{1.45}} & \multicolumn{1}{c}{} & \multicolumn{1}{c}{\textbf{\deleted{Total}}} & \multicolumn{1}{c}{\deleted{26}} & \multicolumn{1}{c}{\deleted{-}} & \multicolumn{1}{c}{\deleted{\textbf{25/25 (100.0\%)}}} & \multicolumn{1}{c}{\deleted{0.64}} & \multicolumn{1}{c}{\deleted{\textbf{60/67 (89.6\%)}}} & \multicolumn{1}{c}{\deleted{1.13}} & \multicolumn{1}{c}{} \\}
\multicolumn{1}{c}{\textbf{\added{Total}}} & \multicolumn{1}{c}{\added{56}} & \multicolumn{1}{c}{\added{-}} & \multicolumn{1}{c}{\added{\textbf{31/35 (88.6\%)}}} & \multicolumn{1}{c}{\added{2.55}} & \multicolumn{1}{c}{\added{\textbf{79/80 (98.8\%)}}} & \multicolumn{1}{c}{\added{1.80}} & \multicolumn{1}{c}{} & \multicolumn{1}{c}{\textbf{\added{Total}}} & \multicolumn{1}{c}{\added{39}} & \multicolumn{1}{c}{\added{-}} & \multicolumn{1}{c}{\added{\textbf{32/32 (100.0\%)}}} & \multicolumn{1}{c}{\added{3.15}} & \multicolumn{1}{c}{\added{\textbf{79/88 (89.8\%)}}} & \multicolumn{1}{c}{\added{1.80}} & \multicolumn{1}{c}{} \\
\bottomrule
\end{tabular}
\\\vspace{.3ex}
\noindent\makebox[\dimexpr\linewidth-1.5cm][r]{\small$^{\ast}$
  Validity ratio is 
  \added{97\% at minimum} for both
  ground truth and \tname in all APIs.}
\end{table*}

%% file: figures/rq1_validity_bars_table.tex
\begin{figure*}[t]
    \centering
    \begin{minipage}{0.58\textwidth} 
        \centering
        \begin{subfigure}[b]{0.48\textwidth}
            \centering
            \includegraphics[width=\linewidth]{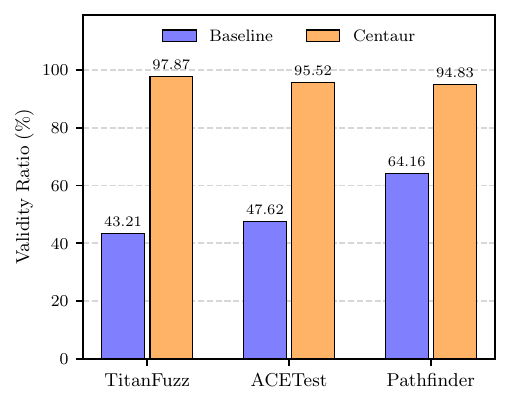} 
            \begin{minipage}{\dimexpr\linewidth+0.8cm}
            \vspace{-3mm}
            \caption{PyTorch validity ratios}
            \label{fig:rq1_validity_pt}
            \end{minipage}
        \end{subfigure}
        \hfill 
        \begin{subfigure}[b]{0.48\textwidth}
            \centering
            \includegraphics[width=\linewidth]{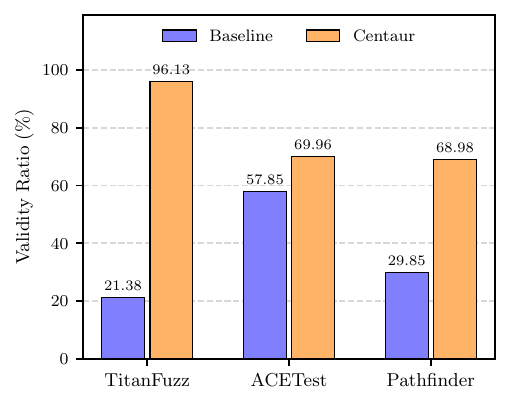}
            \begin{minipage}{\dimexpr\linewidth+0.8cm}
            \vspace{-3mm}
            \caption{TensorFlow validity ratios}
            \label{fig:rq1_validity_tf}
            \end{minipage}
        \end{subfigure}
    \end{minipage}%
    \hfill
    \begin{subfigure}[t]{0.38\textwidth}
    \footnotesize
    \setlength{\tabcolsep}{2pt}  
    \renewcommand{\arraystretch}{0.88}  
    \vspace{-2.2cm}
    \begin{tabular}{ccccc}
    \toprule
        \multirow{2}{*}{\vspace{-1mm}\textbf{Tool}} & \multicolumn{2}{c}{\textbf{Valid Inputs}} & \multicolumn{2}{c}{\textbf{Coverage}}\\
        \noalign{\vskip 0.6mm}\cline{2-3}\cline{4-5}\noalign{\vskip 0.95mm}
    & PyTorch & TensorFlow & PyTorch & TensorFlow \\
        \midrule
        \titanfuzz & \validTitanfuzzPtBase & \validTitanfuzzTfBase & 10,012 & 4,427 \\
        \tname & \validTitanfuzzPtSlate & \validTitanfuzzTfSlate & 10,080 & 4,562 \\
        $\Delta$ & \deltaTitanfuzzPt & \deltaTitanfuzzTf & \textcolor{blue}{$\uparrow$ 68} & \textcolor{blue}{$\uparrow$ 135}\\
    \midrule
    \acetest & \validAcetestPtBase & \validAcetestTfBase & 9,969 & 4,414\\
    \tname & \validAcetestPtSlate & \validAcetestTfSlate & 10,129 & 4,404\\
        $\Delta$ & \deltaAcetestPt & \deltaAcetestTf & \textcolor{blue}{$\uparrow$ 160} & \textcolor{red}{$\downarrow$ 10}\\
    \midrule
    \pathfinder & \validPathfinderPtBase & \validPathfinderTfBase & 3,090 & 1,905\\
    \tname & \validPathfinderPtSlate & \validPathfinderTfSlate & 10,183 & 4,420\\
    $\Delta$ & \deltaPathfinderPt & \deltaPathfinderTf  & \textcolor{blue}{$\uparrow$ 7,093} & \textcolor{blue}{$\uparrow$ 2,515}\\
    \bottomrule
    \end{tabular}	
    \begin{minipage}{\dimexpr\linewidth-0.6cm}
    \vspace{2mm}
    \caption{Numbers of valid inputs and coverage}
    \label{fig:rq1_coverage}
    \end{minipage}
    \end{subfigure}
    \vspace{-0.1in}
    \caption{Comparison of avg. validity ratios, total valid inputs, and avg. branch coverage (three minutes per API). For each pair of tools, we only consider the common APIs they both support.}
    \label{fig:rq1_combined}
\end{figure*}

%% file: evaluation.tex
\vspace{-1ex}
\section{Evaluation}

\noindent{}We pose the following research questions~(\textbf{\RQs}):

\DefMacro{rq-specs}{\RQ{1}:~How effective is the \invariant learning of \tname{}?} 

\DefMacro{rq-comparison}{\RQ{2}:~How does \tname\ compare with the
  \sota in terms of validity rate and coverage?}

\DefMacro{rq-in-the-wild}{\RQ{3}:~How effective is \tname\ in
  detecting new bugs?} 


\begin{packed_itemize}
  \item[] \noindent\textbf{\UseMacro{rq-specs}}\\    
  \item[] \noindent\textbf{\UseMacro{rq-comparison}}\\  
  \item[] \noindent\textbf{\UseMacro{rq-in-the-wild}}\\
\end{packed_itemize}
\vspace{-2ex}

The first question evaluates the quality of \invariants \tname\ learns based on
manually written ground truth. The second question compares \tname\ against
\sota\ fuzzing tools using the standard metrics in the field. The third question
evaluates the ability of \tname\ to reveal deep bugs in \dl\ APIs.

\vspace{-1ex}
\subsection{RQ1: Effectiveness of Constraint Learning}

To evaluate the quality of \tname-learned \invariants,
we select 
15 APIs from PyTorch and TensorFlow each, and manually extract ground truth
constraints. We refer to the API documentation and error messages (\#E) to
understand the input constraints for a given API and build the ground truth. We
randomly selected the APIs under the conditions that
1)~at least 25\% of random inputs are invalid (the average validity ratio is
13.9\% across all APIs),
and that 2)~\tname generates
less than 15 \invariants. The former is to ensure that we can
compare interesting constraints and the latter is to balance manual
effort. To validate the ground truth correctness, we write rules in
our grammar and use the translator to generate Python code.  This
allows us to
check if 1) all valid inputs pass the rules, and 2) the ground truth
enables valid input generation.
We confirm that all the \invariants pass with valid inputs, and improve the low
validity ratio to 
\added{97\% at minimum}. For the same sets of APIs, \tname generates
\added{\totalLeftISDen}
and 
\added{\totalRightISDen} \invariants, respectively. We conduct manual analysis
to evaluate the recall and precision of these constraints against the ground
truth.

Table~\ref{tab:rq1-inv} shows how many of the ground 
truth \invariants \tname covers ($I^{G}$), and how many of \tname-generated 
\invariants are correct ($I^{S}$). Across PyTorch and TensorFlow,
\tname is accurate and precise: it has a recall of 100\% for 
\added{\numDaggeredRecall} APIs and a precision of 100\% for 
\added{\numDaggeredPrecision}
APIs ($\dagger$). \tname covers 
\added{\totalLeftIGNum{} (\totalLeftIGPct)} and 
\added{\totalRightIGNum} (\totalRightIGPct) of the ground truth
\invariants.  For the 
\added{four missed} \invariants, 
we find that 
\added{\tname either removes the constraints as they do not affect validity
ratio, or is unable to generate the rule due to the string literals not
supported in the grammar.}
Overall, we find that 
\added{\totalLeftISPct} (\torch)
and 
\added{\totalRightISPct} (\tf) of the generated \invariants are correct,
demonstrating that \tname can generate high quality constraints for \dl APIs.
Further, we observe that \tname's model generation is efficient: they require
only few seconds to generate models for each API (columns $M^{G}$ and $M^{S}$).

\vspace{-2ex}
\subsection{Comparison with \sota{} DL Library Fuzzers}

\DefMacro{rq-validity}{\RQ{2.1}:~How does \tname\ compare with the
  \sota\ in terms of validity ratio?}

\DefMacro{rq-coverage}{\RQ{2.2}:~How does \tname\ compare with the
  \sota\ in terms of coverage?}

\subsubsection{Answering \UseMacro{rq-validity}}\label{eval:validity}
\label{sec:rq2-val}

Figures~\ref{fig:rq1_validity_pt} and~\ref{fig:rq1_validity_tf} show avg.
validity ratios for \tname and the baselines considering all common
supported APIs. The validity ratio is calculated as the ratio of valid inputs to the total
number of inputs generated by each tool. The results show that \tname\ achieves
a significantly higher validity ratio compared to \sota\ tools across both
libraries. For example, \tname\ achieves a validity ratio of
\added{97.87}\% for \torch, while \titanfuzz only achieves 
\added{43.21}\%. 
\Space{Similarly, for \tf, \tname\ achieves a validity ratio of 
\added{96.13}\%, while
\titanfuzz only achieves 
\added{21.38}\%. }
The closest any approach gets to \tname's
validity ratio is 
\added{\pathfinder's 64.16\% for \torch and \acetest's 57.85\% for
\tf}. It is no surprise that both techniques are constraint-based and they do
much better than \titanfuzz. This result further demonstrates the quality of \tname's
\invariants compared to \acetest~and \pathfinder.

\added{In addition to the overall validity ratio, we also analyze the
statistical significance of the differences between the validity ratio
distributions of \tname\ and each baseline. We use the same standard statistical
tests as in the coverage comparison (\textsection~\ref{sec:rq2-cov}). The
p-values we obtain reject the null hypothesis that the distributions have no
significant statistical difference. The p-values are \tTestPvalueTFVal,
\tTestPvalueACVal, and \tTestPvaluePFVal\ for \titanfuzz, \acetest, and
\pathfinder\ on \torch, respectively. For \tf, the p-values are
\tTestPvalueTFtfVal, \tTestPvalueACtfVal, and \tTestPvaluePFtfVal\ for
\titanfuzz, \acetest, and \pathfinder, respectively. We also calculate the
Cohen's d values to measure the effect size between the distributions. The
effect sizes are very large for all comparisons in \torch, and large to medium
in \tf. Cohen's d values of \cohensDvalueTFVal, \cohensDvalueACVal, and
\cohensDvaluePFVal\ for \torch, and \cohensDvalueTFtfVal, \cohensDvalueACtfVal,
and \cohensDvaluePFtfVal\ for \tf, for \titanfuzz, \acetest, and \pathfinder,
respectively.}

The second and third columns in Figure~\ref{fig:rq1_coverage}
show the numbers of valid inputs generated across all APIs. Centaur
generates significantly more valid inputs than the baselines for
both libraries. For example, Centaur generates 65,479k and 56,248k
inputs, whereas the baselines generate 5,658k and 34k inputs for
PyTorch and TensorFlow, respectively. TitanFuzz makes an LLM
request for every input generation, and the path conditions for
ACETest and Pathfinder enable only one abstract input. In 
contrast, Centaur’s constraints are based on API parameters, which
enable the generation of multiple abstract inputs. In summary, 
Centaur’s constraints are not only accurate—achieving high validity
ratios—but also efficient in enabling input generation.


\subsubsection{Answering \UseMacro{rq-coverage}}\label{sec:rq2-cov}

We collect branch coverage for each \sota\ per API and compare
corresponding distributions.  Table~\ref{tab:num_target_apis} contains
information on the number of APIs on which we ran this evaluation
on \tname\ and the number of APIs that intersect with each comparison
baseline.
The last two columns in Figure~\ref{fig:rq1_coverage}
present the 
\added{average} number of
branches covered by each technique across all APIs using a time budget of 180 seconds.
Comparing \tname with \pathfinder{}, the differences between the 
distributions are statistically significant with
p-values \tTestPvaluePF\ and
\tTestPvaluePFtf\ for \torch\ and \tf\ respectively. The effect size is very
large, with Cohen's $d$ values of \cohensDvaluePF\ and \cohensDvaluePFtf\ for
\torch\ and \tf. \pathfinder{} only focuses on a limited component of the
backend, which prevents it from covering branches in other parts of
the DL libraries that \tname\ can explore. However, for \acetest\
and \titanfuzz, we do not observe statistical significance between the
distributions. The coverage distribution from
\tname\ have a small effect size when compared to \titanfuzz (\cohensDvalueTF\ and
\titanfuzz{} \cohensDvalueTFtf\ for \torch\ and \tf), but no measurable
differences when compared to \acetest. On average, \tname\
demonstrates a slight advantage over \acetest{} and \titanfuzz{},
covering \moreCovThanACETest\ and \moreCovThanTitanfuzz\ more
branches on average per API than
\acetest\ and \titanfuzz, respectively. 




\begin{table}[t!]
\vspace{-2ex}
\centering
  \caption{\added{Summary of r}eported bugs.}\label{tab:bugs}
  \footnotesize
  \setlength{\tabcolsep}{1.5pt}
  \vspace{-3ex}
    \begin{tabular}{lrrrrrrr}\toprule
    \textbf{Library} & \textbf{Submitted} & \textbf{Confirmed} & \textbf{Rejected} & \textbf{Duplicates} & \textbf{Pending} & \textbf{Fixed} \\ \midrule
    \torch & \torchBugsSubmitted & \torchBugsConfirmed & \torchBugsRejected & \torchBugsPreviouslyFixed & \torchBugsPending & \torchBugsFixed \\
    \tf & \tfBugsSubmitted & \tfBugsConfirmed & \tfBugsRejected & \tfBugsPreviouslyFixed & \tfBugsPending & \tfBugsFixed \\ \midrule
    $\Sigma$ & \totalBugsSubmitted & \totalBugsConfirmed & \totalBugsRejected & \totalBugsPreviouslyFixed & \totalBugsPending & \totalBugsFixed \\
    \bottomrule
    \end{tabular}
  \vspace{-1.5ex}
  \end{table}

\begin{table}[!t]\centering
\small
\caption{Categorization of the reported bugs.}
\vspace{-3ex}
\label{tab:bug-categories}
\renewcommand{\arraystretch}{0.8}
\setlength{\tabcolsep}{6pt}
  \begin{tabular}{lrrrr}
  \toprule \textbf{Category} & \textbf{Bugs in \torch} & \textbf{Bugs in \tf}  & $\Sigma$ \\
  \midrule
  Crash & \torchBugsCrash & \tfBugsCrash & \totalBugsCrash \\
  Inconsistent & \torchBugsInconsistent & \tfBugsInconsistent & \totalBugsInconsistent \\
  Overflow & \torchBugsOverflow & \tfBugsOverflow & \totalBugsOverflow \\
  NaN & \torchBugsNan & \tfBugsNan & \totalBugsNan \\
  \midrule $\Sigma$ &\torchBugs & \tfBugs & \totalBugs \\
  \bottomrule
  \end{tabular}
\vspace{1.5ex}
\end{table}

\subsection{RQ3: Detecting New Bugs in \dl\ Libraries}

\lstset{
    aboveskip=-0.3em,
    belowskip=-2.0em
}

\begin{lstlisting}[
  style=custompython,
  caption={Bug in \torch{}~\cite{torch_issue_158154}.},
  float,
  xleftmargin=3ex,
  linewidth=\dimexpr\linewidth-2ex\relax,
  label=lst:issue_158154
]
import torch as pt
input = pt.ones(1, 2, 3, 4)
output = pt.native_channel_shuffle(input, groups=3)
# Floating point exception (core dumped)
\end{lstlisting}

\lstset{
    aboveskip=.65em,
    belowskip=-2.0em
}

\begin{lstlisting}[
  style=custompython,
  caption={Example of a deep bug in \tf{} \cite{tf_issue_97105}.},
  float,
  xleftmargin=3ex,
  linewidth=\dimexpr\linewidth-2ex\relax,
  label=lst:issue_97105
]
import torch as pt, tensorflow as tf
a = tf.random.uniform((1,1,3,1), 0, 1, tf.float32)
depth_radius, alpha, beta = 5, 10, -1
pt_cpu = pt.nn.functional.local_response_norm(
  pt.tensor(a.numpy()), depth_radius, alpha, beta)
# tensor([[[[2.7409],[0.6304],[0.7823]]]])
pt_gpu = pt.nn.functional.local_response_norm(
  pt.tensor(a.numpy()).cuda(), depth_radius, 
  alpha, beta)
# tensor([[[[2.7409],[0.6304],[0.7823]]]])
with tf.device('/cpu:0'):
  tf_cpu = tf.nn.local_response_normalization(
  a, depth_radius, alpha, beta)
# tf.Tensor([[[[9.8576],[1.3555],[1.8605]]]])
with tf.device('/gpu:0'):
  tf_gpu = tf.nn.local_response_normalization(
  a, depth_radius, alpha, beta)  
# InvalidArgumentError: {{function_node... 
# cuDNN requires beta >= 0.01, got: -1 [Op:LRN]
\end{lstlisting}

Table \ref{tab:bugs} summarizes bugs we report in \torch\ and \tf. So far, we
found \totalBugs{} bugs in total, of which \totalBugsConfirmed{} were confirmed.
We relied on crash and differential oracles (\ie{}, the output difference
between CPU and GPU) to find bugs.  Results demonstrate the effectiveness of
\tname\ in finding bugs in \dl\ libraries.

\tname\ was successful in finding deep bugs in
\torch\ and \tf.
The issue \CodeIn{\# 158154}~\cite{torch_issue_158154} shows one of such cases
in the API
\texttt{torch.native\_channel\_shuffle}~(Listing~\ref{lst:issue_158154}). The
bug is caused by passing a value to the \CodeIn{groups} parameter that is
greater than the size of the second dimension of the input tensor. The input
appears valid according to the API documentation.  Execution of the test causes
a crash causes by a floating point exception. Our artifact~\cite{ourrepo}
includes the entire list of bugs we reported.

Issue \CodeIn{\# 97105}~\cite{tf_issue_97105}~(Listing~\ref{lst:issue_97105})
shows another example. 
The issue demonstrates a discrepancy where a negative \CodeIn{beta} value passed
on CPU but triggered a cuDNN-related~\cite{chetlur2014cudnn}
\CodeIn{InvalidArgumentError} on \tf{} GPU. Since \torch{} also uses the cuDNN
backend yet yielded consistent, error-free results across devices, we concluded
the fault lies in \tf's implementation rather than the underlying library and
reported the bug.


We note that some bug reports that developers rejected are not false positives.
One such example is issue \CodeIn{\# 158208} \cite{torch_issue_158208} in
\torch.
We report a difference of
$1.0737e+09$ between the CPU and GPU outputs. 
The developers rejected the bug report citing expected behavior due to floating
point arithmetic differences between numerical libraries like LAPACK
\cite{anderson1999lapack} and cuSOLVER \cite{nvidia_cusolver}.

Table~\ref{tab:bug-categories} shows the categories of the bugs we report with
\tname. We find \totalBugsCrash{} bugs in \torch\ and \tf that caused the
program to crash by raising a signal (e.g. Floating Point Exception, Aborted
etc.). These bugs are generally more critical since they can crash existing
AI-based applications without any meaningful logs to help developers debug. We
also find a bug that triggers \CodeIn{NaN} values in the output of the API.
This is the only bug that was fixed by the developers at the time of writing.
\Space{Note that most of the bugs we report (\totalBugsInconsistent{}) are caused by
CPU-GPU inconsistencies. It is worth noting that all of the rejected
bugs were of this kind. This shows a practical weakness of the oracle;
it can not differentiate between failures caused by precision errors
and actual bugs. Even after manual inspection, it is difficult to
filter these out without developer feedback or expert domain
knowledge. To mitigate this, we tried different levels
of \CodeIn{tolerance} i.e., the maximum allowed difference between the
CPU and GPU outputs. We could not converge on a single value that
would mitigate the issue, so we opted to use a value of $0.01$
following other \sota{} implementations.} We also
found \totalBugsOverflow{} overflow bugs. These cases are typically
considered less critical and can be handled by revising input
validation checkers in code.


%% file: discussion.tex
\section{Discussion}\label{discussion}
\input{figures/coverage_plot.tex}
\mypara{Generalizability}
To evaluate \tname's generalizability across different LLMs, we choose four
additional popular LLMs from different families: two closed source models (GPT-5
and Claude Sonnet 4.5) and two open source models (Gemma 3 27B and
Qwen3-Coder-30B).
We randomly sample 100 APIs from \torch\ and 100 APIs from 
\tf\ and apply \tname\ with each LLM to keep the experiment cost-effective.
\autoref{tab:llm} shows
the number of APIs successfully executed, avg. branch coverage, and avg.
validity ratio. For comparison, numbers in parentheses show results for the
default LLM, Gemini 2.0 Flash. Results show that Gemini 2.0 Flash outperforms
other LLMs in terms of successful execution and that all LLMs but
Qwen3-Coder-30B perform well considering the metrics.
We find that while \tname is compatible with multiple LLMs, the
choice of LLM has a moderate impact on effectiveness.
\input{tables/llm.tex}

\mypara{Ablation Study} 
To evaluate necessity of each component of \tname, we conduct an ablation study
that systematically removes one component and evaluates impact on performance.
We select three components of rule generation without which the technique still
produces results: (1) feedback to the LLM on semantic validity of the rules (w/o
Feedback), (2) supplemental information on constraints such as documentation,
error messages (w/o DocErr),  and (3) example rules to include with the prompt
for rule generation (w/o ExRules). We also remove all three together to measure
the effect of a technique where rule generation is performed without any
assistance (w/o All). We randomly sample 100 APIs each from \torch\ and \tf\ to
run the ablation configurations on these apis. \autoref{tab:ablation} shows that
each of these components is essential for \tname\ considering the average
reduction in measurements of ``\# APIs'' and ``Coverage''.
\input{tables/ablation.tex}

\mypara{Correctness of Rule Refinement} 
\tname's refinement based on validity ratio may potentially
discard true constraints. To check its correctness, we select five widely 
used APIs: \CodeIn{torch.argmax}, \CodeIn{torch.nn.functional.alpha\_dropout}, 
\CodeIn{torch.nn.ReLU}, \CodeIn{tf.math.sigmoid}, and \CodeIn{tf.image.transpose},
and manually validate the discarded constraints. In total, \tname 
generated 141 constraints for all five APIs and then discarded 119 among 
them. Our manual analysis confirms that all 119 are correctly excluded and
fall into three categories: (1) duplicates, (2) default constraints 
provided by Centaur (e.g., parameter types), and (3) incorrect constraints.

\mypara{Time and Monetary Costs}
\tname uses LLM to generate API signatures, valid inputs, and candidate
rules. To compute the time and monetary costs of using LLM, we run \tname's
default model (Gemini 2.0 Flash) for the 100 APIs used in ablation. 
For signature and input generation, \tname takes 64.28 and 79.25 seconds, and
uses 94k and 61k tokens on average per API for \torch\ and \tf,
respectively. This costs an average of \$0.013 and \$0.009 per API for
\torch\ and \tf. 
Similarly, for rule generation, time limit is
configurable. We use one minute per API, and \tname generates 93.37 rules on
average during the time limit
(\textsection~\ref{sec:rule-generation-pseudocode}). \tname uses 2,872.5k and
2,234.1k tokens on average per API, which costs an average of \$0.30 and \$0.23
per API for \torch\ and \tf, respectively. The results show that both
time and monetary costs of using LLM are nominal. More importantly, the
generation is a one-time offline process, and the LLM's outputs can be reused
later for fuzzing.
 

\mypara{\added{Selection of time budget for fuzzing}}
\added{We chose 180s as our time budget to balance the cost of running a total
of 1,034 APIs across all baselines. \tname\ can generate tens of millions of inputs
in 180 seconds, allowing it to achieve high coverage in a short period. In
contrast, in the same time frame, the other approaches generate 100x or 1000x
fewer (and less diverse) inputs, and hence achieve lower coverage. 
For further validation, we calculated the cumulative coverage achieved by
\tname\ vs \sota\ tools over a longer time budget of 10 minutes.
\autoref{fig:cov_trend} shows the coverage trend with a snapshot taken every
minute. The results show that \tname\ outperforms all \sota{} tools over both
\torch\ and \tf\ on completion except \titanfuzz{} on \tf{}. On manual analysis,
we found that \titanfuzz{} implicitly knows about hardware optimized
kernel-specific memory format propagation rules for certain APIs in \tf{} due to
its usage of LLMs. For example, in oneDNN~\cite{li2024onednn}, if
\CodeIn{dtype=int8}; then the "channel" dimension of a \CodeIn{NCHW} formatted
tensor is expected to be aligned to 16 bytes~\cite{onednn-mem}. Such
optimization-specific rules are not captured by \tname's inferred input
invariants, leading to lower coverage. } 



%% file: figures/coverage_plot.tex
\begin{figure}[t]
    \vspace{-.2ex}
    \centering
    
    
    
    
    \includegraphics[width=\columnwidth]{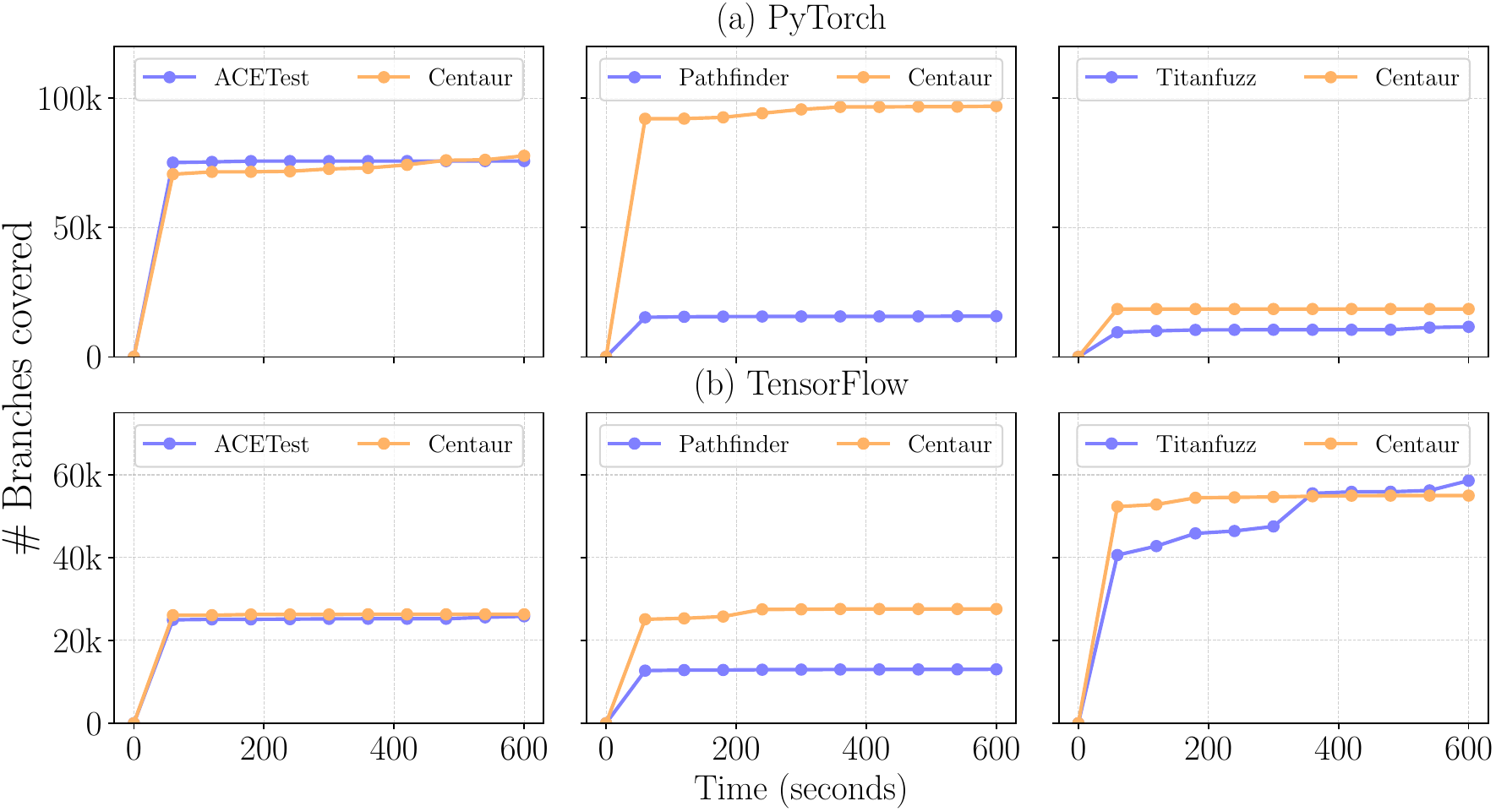}
    \vspace{-5ex}
    \caption{Coverage trend over 10 minutes for \tname{} vs \sota{} tools}
    \label{fig:cov_trend}
    \vspace{-2ex}
\end{figure}

%% file: tables/llm.tex
\begin{table}[t!]
\centering
\begin{minipage}{0.98\columnwidth}
\centering
  \caption{Performance of \tname\ using different LLMs.}\vspace{-2.5ex}
  \label{tab:llm}
  \resizebox{\columnwidth}{!}{%
  \setlength{\tabcolsep}{6pt}
    \begin{tabular}{p{1.5cm}lccc}
    \toprule
    \textbf{Library} & \textbf{Model} & \textbf{\# APIs} & \textbf{Coverage} & \textbf{Validity Ratio} \\
    \midrule

    \multirow{4}{*}{\textsc{PyTorch}}
      & GPT-5               & 79 {\scriptsize (88)}        & 9928.40 {\scriptsize (9910.34)}   & 96.39\% {\scriptsize (95.62\%)} \\
      & Claude Sonnet 4.5   & 68 {\scriptsize (88)}        & 9946.51 {\scriptsize (9922.33)}   & 96.58\% {\scriptsize (96.62\%)} \\
      & Gemma 3 27B         & 70 {\scriptsize (88)}        & 9968.88 {\scriptsize (10001.22)}  & 96.70\% {\scriptsize (96.72\%)} \\
      & Qwen3-Coder-30B     & 39 {\scriptsize (88)}        & 9950.06 {\scriptsize (9965.36)}   & 96.97\% {\scriptsize (96.97\%)} \\
    \midrule

    \multirow{4}{*}{\textsc{TensorFlow}}
      & GPT-5               & 54 {\scriptsize (93)}        & 4600.85 {\scriptsize (4509.09)}   & 93.64\% {\scriptsize (92.94\%)} \\
      & Claude Sonnet 4.5   & 43 {\scriptsize (93)}        & 4410.81 {\scriptsize (4548.11)}   & 99.93\% {\scriptsize (95.39\%)} \\
      & Gemma 3 27B         & 72 {\scriptsize (93)}        & 4339.52 {\scriptsize (4411.10)}   & 92.52\% {\scriptsize (94.20\%)} \\
      & Qwen3-Coder-30B     & 33 {\scriptsize (93)}        & 4496.65 {\scriptsize (4482.26)}   & 96.82\% {\scriptsize (96.79\%)} \\
    \bottomrule\\[-2pt]
    \end{tabular}%
  }
\end{minipage}
\vspace{-0.55cm}
\end{table}

%% file: tables/ablation.tex
\begin{table}[t!]
\vspace{-.15ex}  
\begin{minipage}{0.98\columnwidth}
\centering
    \caption{Results of ablation study.}\vspace{-2.5ex}
    \label{tab:ablation}
    \resizebox{\columnwidth}{!}{%
\begingroup
      \renewcommand{\arraystretch}{0.9}
      \setlength{\tabcolsep}{10pt}
      \begin{tabular}{p{1.5cm}lccc}
      \toprule
      \textbf{Library} & \textbf{Ablation} & \# \textbf{APIs} & \textbf{Coverage} & \textbf{Validity Ratio} \\
      \midrule
\multirow[c]{5}{*}{\textsc{PyTorch}}
        & w/o DocErr & 74/100 & 9,947.65 & 94.57\% \\
        & w/o Feedback & 77/100 & 9,954.31 & 94.79\% \\        
        & w/o ExRules & 81/100 & 9,978.32 & 97.50\% \\
        & w/o All & 82/100 & 9,948.33 & 95.18\% \\
\cmidrule{2-5}
\addlinespace[-.2ex]	         
& \textbf{\tname} & 88/100 & 9,983.99 & 96.48\% \\
      \midrule

\multirow[c]{5}{*}{\textsc{TensorFlow}}
        & w/o DocErr & 43/100 & 4,379.47 & 92.21\% \\
        & w/o Feedback & 55/100 & 4,298.61 & 95.40\% \\
        & w/o ExRules & 45/100 & 4,510.29 & 97.76\% \\
        & w/o All & 48/100 & 4,341.12 & 89.93\% \\
\cmidrule{2-5}
\addlinespace[-.2ex]	         
& \textbf{\tname} & 93/100 & 4,462.66 & 94.24\% \\
      \bottomrule
      \end{tabular}%
\endgroup
    }
\end{minipage}
\vspace{-4ex}
\end{table}

%% file: validity.tex
\section{Threats to Validity}
\mypara{\added{External Validity}} 
\added{The generality of \tname\ is limited by our selection of APIs, the 
expressiveness of our grammar, and the sample size used in the
qualitative study. Considering API selection, we chose APIs from \torch\ and \tf\ based on their popularity
and usage in prior work~\cite{Shi_ETAL_ISSTA23,Deng_ETAL_ISSTA23,Kim2025LightweightConcolic}.
It is also worth noting that our tool is extensible; by running the offline phases of \tname\ with
scripts provided in our replication package~\cite{ourrepo}, users can easily apply \tname\ to
other APIs in these libraries.
Considering our grammar, there may be rules that it cannot express, 
e.g., when a constraint is conditioned on an optional
argument. One example is: ``if an argument for the \CodeIn{axis} parameter is
given, the value must be a valid dimension of the \CodeIn{input} tensor'' 
concerning the API \CodeIn{tf.reduce\_sum}. 
We designed our grammar to express a variety of 
constructs and cover invariants of a large number of APIs.
Also, our grammar is based on 
the Lark format~\cite{LarkParser} and can be easily extended. 
Considering the qualitative analysis of learned constraints, the small
sample size of APIs and potential errors may have impacted precision and recall. To mitigate this threat, we carefully selected
the APIs to cover different functionalities (e.g., \CodeIn{nn}, \CodeIn{math},
\CodeIn{linalg}, and \CodeIn{experimental} modules) and only
considered the APIs with clear error messages to facilitate ground
truth definition.}

\mypara{Internal Validity} \tname's oracles may report false positive bugs. So,
we conduct rigorous manual validation for every reported bug and create
minimized test cases to reproduce bugs.
Using a small set of inputs can cause wrong candidate rules less 
effectively pruned. To mitigate this, we first obtain a small number of 
inputs using an LLM and apply nine types of carefully-designed mutations to 
enrich the set. In the end, we use 117 valid inputs on average per API.
Lastly, the LLM's randomness may lead to different
results for different runs. However, we believe that its impact is minimized
because \tname (1) uses an iterative approach to generate rules, and (2) applies
the approach across all APIs of both libraries. Also, during implementation, 
we ran the rule generation multiple times and did not observe significant 
differences in quality.


%% file: related.tex
\section{Related Work}

\mypara{Model-level Testing of Deep Learning Libraries}
CRADLE~\cite{Phan_ETAL_ICSE2019} was one of the first approaches for testing DL
libraries using differential testing by comparing outputs from different
execution backends (e.g., CPU vs. GPU) within the same library. More recently,
Li et al.~\cite{li2024dllens} used Large Language Models (LLMs) to translate
APIs, enabling differential testing between different DL libraries. Similarly,
many other approaches have been proposed to test DL libraries using differential
testing at the model
level~\cite{AUDEE_ASE20,LEMON_FSE20,MUFFIN_ICSE22,Liu_2023,Liu_ETAL_ICSE23}.
Model-level testing is more limited because the number of APIs that can
construct a model by being used together is a small subset of all APIs 
available in the DL libraries. Thus, \tname\ focuses on API-Level testing
to cover a larger set of APIs.

\mypara{API-level Testing of Deep Learning Libraries} API-level
testing~\cite{Wei2022,Deng_ETAL_FSE22,Christou_ETAL_USENIX23,Deng_ETAL_ISSTA23}
is another complementary approach to model-level testing. It works by generating
inputs only for the given API and checking correctness of the outputs via
differential testing and checking for crashes or exceptions. Because API-level
specifications are incomplete, these tools employ various techniques to generate
valid inputs. FreeFuzz~\cite{Wei2022} addresses this by mining open-source
projects to infer API argument types and create valid inputs for mutation.
DocTer~\cite{Xie_ETAL_ISSTA22} complements this by extracting API constraints
(e.g., types, tensor shapes) directly from technical documentation. To improve
testing efficiency, DeepREL~\cite{Deng_ETAL_FSE22} identifies APIs with similar
behaviors to share test inputs across them, increasing coverage. More recent
approaches leverage LLMs. In contrast to these approaches,
IvySyn~\cite{Christou_ETAL_USENIX23} leverages this with a type-aware, black-box
mutator and can map low-level C++ crashes back to high-level Python API calls.
Unlike \tname{} these approaches do not rely on concrete specifications and are
hence prone to generating many invalid inputs. While DocTer and DeepREL also use
constraints, they are much simpler and do not capture the relational constraints
between API parameters that \tname\ can infer.

%
Pathfinder~\cite{Kim2025LightweightConcolic} is a gray-box fuzzer that uses program
synthesis to approximate path conditions, guiding input generation toward new
code paths. ACETest~\cite{Shi_ETAL_ISSTA23} extracts precise API input
constraints by analyzing the library's internal validation code, tracing
backward from error-handling functions. It then uses an SMT solver to generate
valid inputs that satisfy these constraints. While \tname{} also uses SMT
solving, \added{it does so by analyzing inputs and outputs; it does not
require analyzing code and path constraints, which can be
computationally expensive.}


\mypara{LLM-based Testing Approaches}  Titanfuzz~\cite{Deng_ETAL_ISSTA23} uses
LLMs to generate and mutate seed programs for fuzzing, while
FuzzGPT~\cite{deng2024large} generates test code from bug reports and existing
code snippets. Unlike these approaches, \tname{} uses LLMs to generate API
constraints instead of inputs for fuzzing, enabling us to develop a \emph{input
generator} that generates valid and diverse inputs systematically.

\mypara{Invariant Learning} \tname is inspired by
Daikon~\cite{ErnstPGMPTX2007,Ernst2001Daikon, Ernst1999DaikonICSE}, a dynamic
detection technique to infer likely invariants.
Daikon infers invariants from a set of pre-defined patterns and by using a fixed set of execution
traces. In contrast, \tname uses a domain-specific grammar and validates
candidate rules using dynamically generated inputs. Conceptually, this enables learning more
precise, relevant, and complete constraints. Also, unlike Daikon, \tname refines
the constraints. DySy~\cite{Csallner2008DySy}
leverages dynamic symbolic execution and infers 
invariants based on path conditions. 
But, \tname's invariants result in higher validity ratio than those based 
on path conditions (Sec.~\ref{eval:validity}).
ISLearn~\cite{Steinhofel2022InputInvariants} is another Daikon-style invariant
inference approach for system inputs. However, in constrast to \tname{}, ISLearn
needs manual template definition and cannot express complex constraints of DL
library APIs. Lastly, none of the approaches use LLMs -- an essential source of
implicit invariants.

%




%% file: conclusion.tex
\section{Conclusion}
\label{sec:conclusion}
\looseness=-1 We presented \tname{} -- the first neurosymbolic
technique that combines LLMs with SMT constraint solving to learn API
input constraints and generate valid test inputs. We showed that
\tname{} generates high quality constraints for numerous PyTorch and
TensorFlow APIs and achieves higher coverage and validity ratio
compared to \sota{} DL fuzzers. Finally, we showed that \tname{} can
find new bugs in \torch{} and \tf{} libraries. We believe that such
neurosymbolic techniques for constraint learning are quite promising
and general. Such approaches can be further enhanced to find deeper
and more diverse set of bugs not only in \dl\ libraries but also other
domains when input specifications are sparse.
Our replication package is publicly available 
at \RepoURL.
